\documentclass[aps,prb,twocolumn,superscriptaddress,floatfix,longbibliography,citeautoscript]{revtex4-2}

\usepackage{amsmath,amssymb} % math symbols
\usepackage{bm} % bold math font
\usepackage{graphicx} % for figures
\usepackage{comment} % allows block comments

\usepackage{hyperref}
\hypersetup{
colorlinks=true,
urlcolor= blue,
citecolor=blue,
linkcolor= blue,
}

\newcommand{\cmse}{Department of Materials Science \& Engineering, Cornell University, Ithaca, New York 14853, USA}
\newcommand{\kav}{Kavli Institute at Cornell for Nanoscale Science, Cornell University, Ithaca, New York 14853, USA}
\newcommand{\caep}{School of Applied and Engineering Physics, Cornell University, Ithaca, New York 14853, USA}
\newcommand{\CPfS}{Max Planck Institute for Chemical Physics of Solids, Dresden 01187, Germany}
\newcommand{\ummse}{Department of Materials Science \& Engineering, University of Michigan, Ann Arbor, Michigan 48103, USA}
\newcommand{\rowland}{Rowland Institute at Harvard, Harvard University, Cambridge, Massachusetts 02142, USA}

\newcommand{\nrel}{National Renewable Energy Laboratory, Golden, Colorado 80401, USA}

\newcommand{\mytitle}{Quantitative approaches for multi-scale structural analysis with atomic resolution electron microscopy}

\newcommand{\SBMO}{SmBaMn$_2$O$_6$ }
\newcommand{\SRO}{Sr$_2$RuO$_4$ }
\newcommand{\SLAO}{SrLaAlO$_4$ }

\begin{document}

\title{\mytitle}

\author{Noah Schnitzer}
\email{nis29@cornell.edu}
\affiliation{\cmse}
\affiliation{\kav}

\author{Lopa Bhatt}
\affiliation{\caep}

\author{Ismail El Baggari}
\affiliation{\rowland}

\author{Robert Hovden}
\affiliation{\ummse}

\author{Benjamin H. Savitzky}
\email{ben.savitzky@gmail.com}
\affiliation{\caep}

\author{Michelle A. Smeaton}
\affiliation{\nrel}

\author{Berit H. Goodge}
\email{Berit.Goodge@cpfs.mpg.de}
\affiliation{\CPfS}

\date{\today}

\begin{abstract}
Atomic-resolution imaging with scanning transmission electron microscopy is a powerful tool for characterizing the nanoscale structure of materials, in particular features such as defects, local strains, and symmetry-breaking distortions. 
In addition to advanced instrumentation, the effectiveness of the technique depends on computational image analysis to extract meaningful features from complex datasets recorded in experiments, which can be complicated by the presence of noise and artifacts, small or overlapping features, and the need to scale analysis over large representative areas. 
Here, we present image analysis approaches which synergize real and reciprocal space information to efficiently and reliably obtain meaningful structural information with picometer scale precision across hundreds of nanometers of material from atomic-resolution electron microscope images. 
Damping superstructure peaks in reciprocal space allows symmetry-breaking structural distortions to be disentangled from other sources of inhomogeneity and measured with high precision. 
Real space fitting of the wave-like signals resulting from Fourier filtering enables absolute quantification of lattice parameter variations and strain, as well as the uncertainty associated with these measurements. 
Implementations of these algorithms are made available as an open source Python package.
\end{abstract}

\maketitle

\section{\label{sec:Intro}Introduction}
Since the development of aberration correction, atomic-resolution imaging with scanning transmission electron microscopy (STEM) has been increasingly depended upon as an invaluable part of the materials characterization tool set. 
In addition to sub-Angstrom spatial resolution, atomic column positions can be measured with picometer-scale precision, and sensitivity to composition or local lattice strain can be tuned with high- and low-angle collection geometries \cite{yankovich2014picometre, phillips2012atomic}. 
More recently, broad \textit{in situ} capacity, including high resolution imaging at variable temperatures and under applied external fields, makes possible direct correlation of material properties with atomic-scale structural changes, while developing techniques such as phase retrieval with electron ptychography promise to extend these capabilities even further \cite{bianco2021atomic,nukala2021reversible,chen2021electron,smeaton2024tutorial}. 

In the STEM, a condensed (typically $<$1 Å) electron beam is scanned across a thin crystalline sample oriented along a high-symmetry axis to resolve columns of atoms in projection along the beam direction. 
The resulting atomic-resolution images contain rich information about spatial variations in material composition, symmetry, and defects encoded in the image contrast. 
These images are exceptionally useful in studying structural features such as local fields including polarization and strain \cite{jia2007unit,hytch2003measurement}, broken symmetries including those associated with charge or chemical order \cite{savitzky2017bending,abe2003atomic}, as well as the presence of various crystalline phases, domains, defects, and other sources of disorder \cite{browning2001application}. 
A key challenge, however, is disentangling the particular features of interest from the large amount of structural information present in STEM images, as well as in the presence of measurement noise and imaging artifacts. 

\begin{figure}[ht]
    \includegraphics[clip=true,width=\columnwidth]{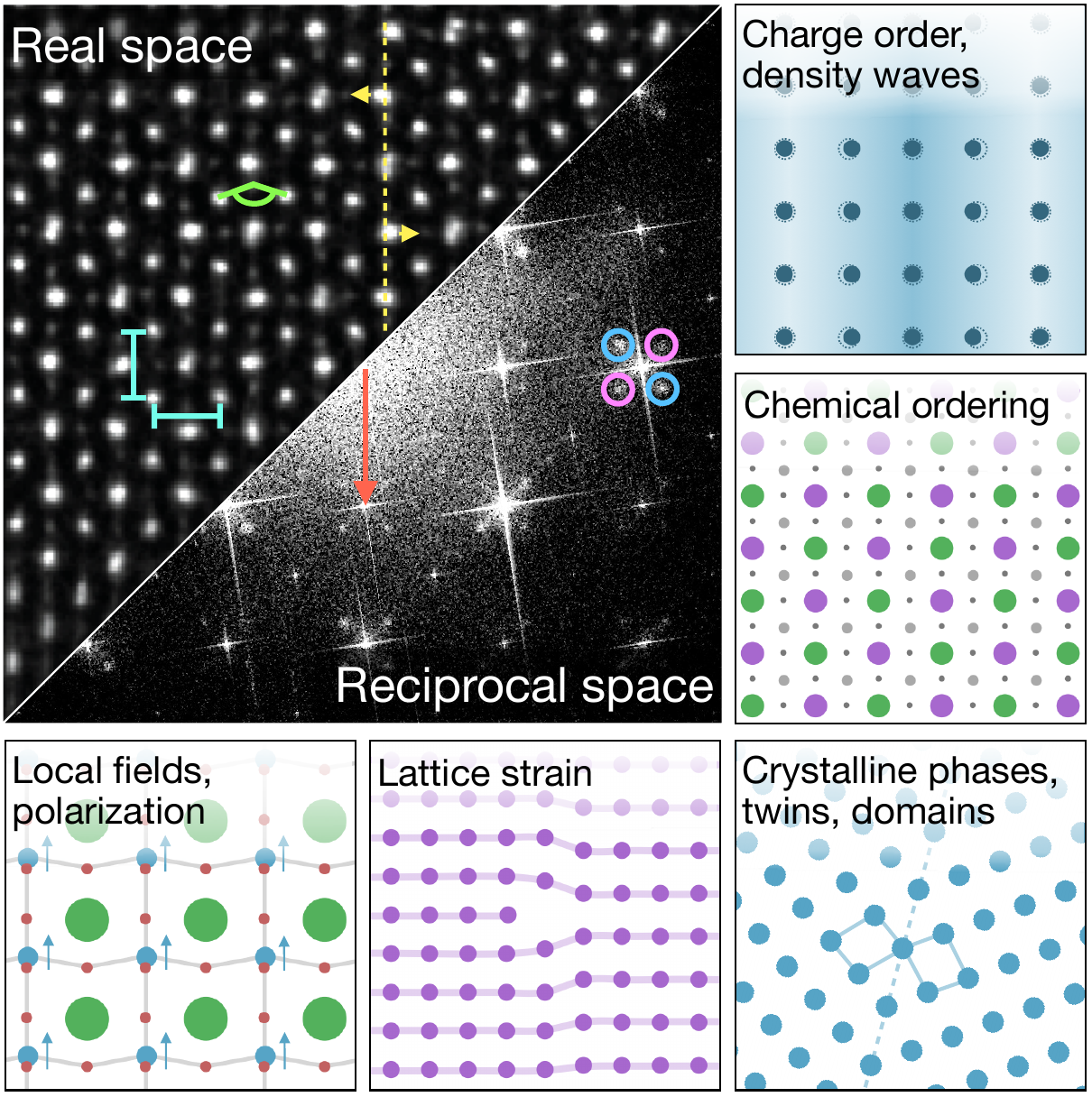}
    \caption{Overview of atomic resolution STEM analysis. Center: a real space STEM image annotated with interplanar spacings (cyan), inter-column angles (green), and displacements relative to high symmetry positions (yellow); and the Fourier transform of another image with Bragg peaks (orange arrow) and satellite peaks associated with two coexisting superlattice structures (cyan, magenta) marked. Surrounding: cartoons illustrating features of interest which can be characterized with atomic resolution STEM imaging.}
     \label{fig:F1}
\end{figure}

A variety of analysis approaches have been developed to handle the many distinct structural features of interest (e.g., inter-planar spacings,  inter-atomic column angles, magnitudes of displacements, etc., as depicted in Figure \ref{fig:F1}).
These approaches range from simple operations, such as extracting a profile of image intensity along a lattice plane, to complex algorithms and associated specialized software packages such as pycroscopy \cite{somnath2019usidpycroscopyopen,mukherjee2020stemtool}, Atomap \cite{nord2017atomap}, TopoTEM \cite{o2022topotem}, motif-learn \cite{dan2022learning,dan2024symmetry}, and TEMMETA \cite{cautaertsDin14970TEMMETATEMMETA2021}. While some analyses can be accomplished entirely in real space, other approaches benefit from working in reciprocal space, which can be accessed by taking the Fourier transform of the STEM image.
Reciprocal space offers a convenient basis for measuring symmetries and the spacing and orientation of lattice planes, while the separation of features based on spatial frequency can be taken advantage of to map specific periodicities back to real space \cite{el2021charge, hytchGeometricPhaseAnalysis1997, hytchAnalysisVariationsStructure1997,savitzky2017bending}. 
For instance, spatial variations in an inter-planar spacing might be measured in real space by calculating the distances between sets of atomic columns, or with Fourier based geometric phase analysis (GPA) \cite{jia2007unit,hytchGeometricPhaseAnalysis1997,hytch2003measurement}. Working in real space and reciprocal space each have advantages and disadvantages in terms of measurement precision and resolution, as well as different levels of susceptibility to noise, computational efficiency, and other trade-offs. Here, we present two approaches that synergize real and reciprocal space analysis taking advantage of the distinct benefits of working in both spaces.

These image processing algorithms address two key STEM image analysis problems: (1) measuring translational symmetry-breaking structural distortions, and (2) quantifying variations in lattice parameters and local strains. 
Our Fourier damping approach for characterizing symmetry breaking can be applied to a wide range of distortions including periodic lattice displacements, octahedral tilting, and chemical or vacancy orders, is robust in the presence of defects, strain and inhomogeneity, and can disentangle coexisting distortions \cite{savitzky2017bending}. 
Meanwhile lattice parameter measurements based on real-space wave fitting to Fourier filtered signals offer interpretable quantification of inter-planar spacings and other parameters on an absolute scale, as well as an understanding of the associated uncertainty in the measurements \cite{smeaton2021mapping}.
While here we focus on high-angle annular dark-field (HAADF) STEM imaging, these approaches are also applicable to other STEM modalities, ptychographic reconstructions, and conventional high-resolution TEM analysis, and can be extended to other high-resolution real space characterization techniques such as scanning probe microscopies.
The algorithms and a suite of atomic-resolution image analysis tools are available in the open source \texttt{kemstem} Python package at \url{https://github.com/noahschnitzer/kemstem}. 

 \begin{figure*}[ht]
    \includegraphics[clip=true,width=\textwidth]{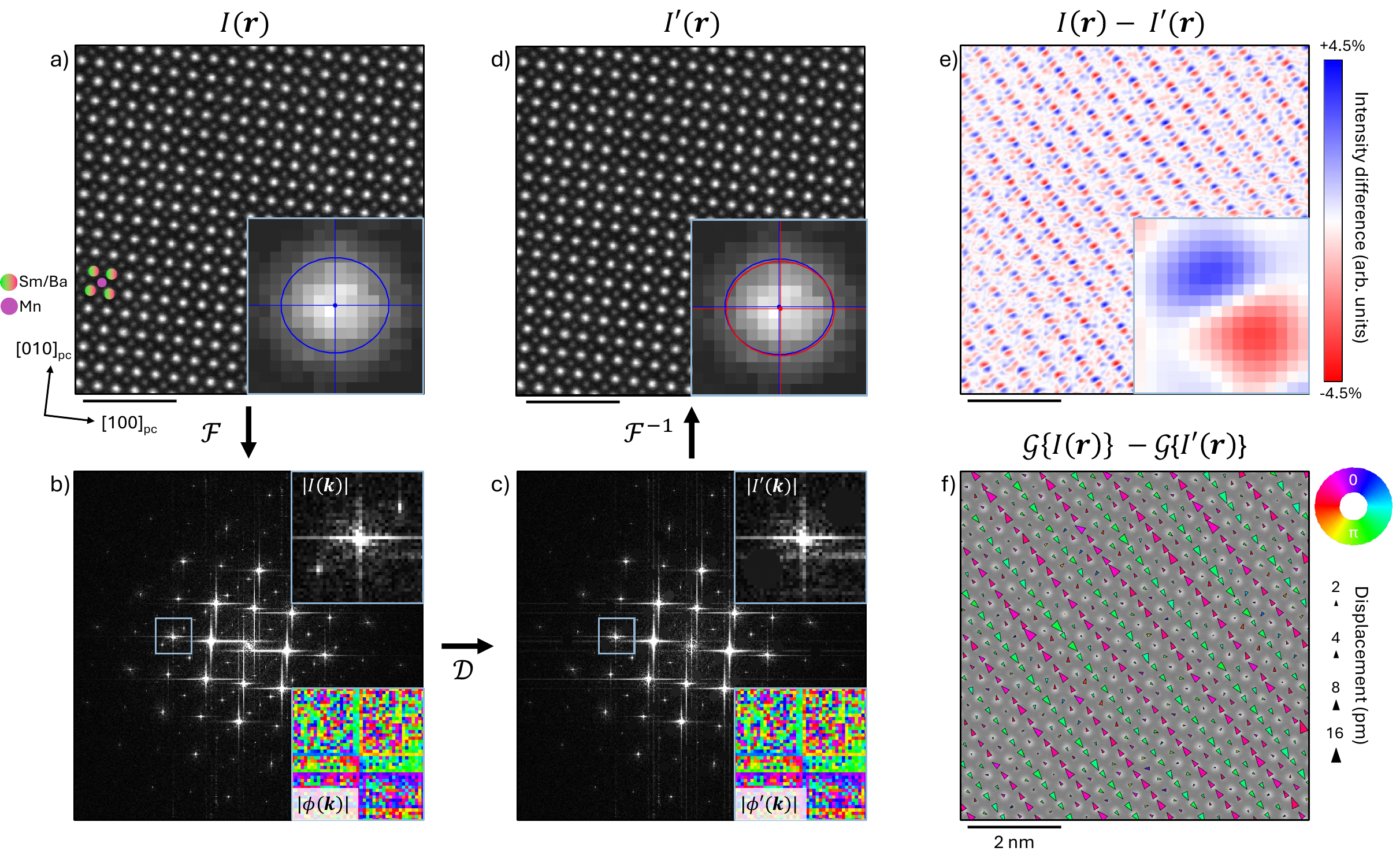}
    \caption{Using Fourier damping to measure and visualize periodic lattice displacements. 
    a) A cropped region of an atomic resolution HAADF-STEM image of a \SBMO single crystal on the [001] zone axis. The inset shows one atomic column, with its center measured from a 2D Gaussian fit marked in blue.
    b) Fourier transform ($\mathcal{F}$) of the image ($I(\textbf{r})$), top-right inset shows the (200) Bragg peak and neighboring superlattice peaks, lower-right inset shows the phase from the same area.
    c) Fourier transform after damping ($\mathcal{D}$) of the superlattice peaks. Insets show the suppression of their magnitude to the background level, while the phase remains unchanged.
    d) Inverse transform after damping ($I('\textbf{r})$), resulting in an undistorted reference image. While visually almost indiscernible from (a), the suppression of the periodic lattice displacements can be seen in the shift of the center of the column in the inset (red) relative to its position in the original image (blue).
    e) Difference between the original (a) and Fourier damped (d) images showing intensity changes associated with the structural distortion. Inset centered on the same atomic column shows a characteristic dumbbell pattern, as the shift results in reduced intensity on one side of the column (red) and increased intensity on the other (blue). 
    f) Atomic displacements are visualized with arrows denoting the shifts between the original and reference positions, measured by Gaussian fitting ($\mathcal{G}$) of each atomic column. The shifts are on the order of picometers - smaller than a single pixel of the STEM image - so the arrows are scaled for visibility, and colored according to the angle of the displacement vector. 
    }
    \label{fig:F2}
\end{figure*}

\section{\label{sec:damp}Mapping symmetry breaking with Fourier damping}
Material functionality and properties of interest are directly tied to crystal symmetry and symmetry breaking structural distortions \cite{newnham2004symmetry}. 
For instance, polar distortions which break inversion symmetry are necessary for ferroelectricity, coupling of charge and orbital order to the lattice break rotation and translational symmetries, and manipulating mirror and inversion symmetries can tune photonic properties \cite{newnham2004symmetry, kivelson2003detect,jacobsen2006strained, fedotov2006asymmetric}. 
The symmetry group of a distorted structure will be a subgroup of a ``parent'' undistorted higher symmetry structure, so the natural basis in which to characterize a broken symmetry is with reference to the parent structure. 
For STEM image analysis, this poses the problem of constructing a high symmetry reference for comparison with the experimental measurement -- generally not a trivial task. 
In some cases, a directly measurable feature of interest such as a change in inter-column distances or angles can be used as an \textit{ad hoc} surrogate, but in many cases no such proxy is available. 
We will show that for translational symmetry breaking distortions – those with a non-zero wavevector - damping the associated peaks in reciprocal space is a systematically applicable method for generating a high symmetry reference to enable the mapping of structural distortions in STEM images \cite{savitzky2017bending,savitzkyComputationalAnalysisOrder2018}.

Breaking translational symmetry forms a superlattice with a supercell larger than the parent unit cell, and a corresponding reciprocal space lattice with smaller reciprocal lattice wavevectors than the parent. 
Thus, in the Fourier transform of a STEM image, the structural information associated with a translational symmetry breaking distortion will be localized to frequencies at the reciprocal lattice points associated with the superstructure. 
A portion of these superlattice points will be distinct from the reciprocal lattice points of the parent, forming satellite peaks at linear combinations of the distortion wavevectors around the parent lattice points (e.g. as marked in cyan and pink in Fig. \ref{fig:F1}) creating a convenient partitioning in reciprocal space between the distorted and parent structures.

Taking advantage of this feature separation in reciprocal space, a high-symmetry reference can be generated from an experimental image by damping the amplitude of the superlattice peaks in its Fourier transform to the background level - without altering the phase - and taking the inverse Fourier transform to change back to a real space basis. 
In the resulting false image, the symmetry breaking distortion will be suppressed, resembling the undistorted parent structure. 
\textcolor{black}{
This can be understood in terms of the Fourier shift theorem: damping the peak amplitudes in Fourier space re-weights the image's phase information, resulting in changes in image contrast and shifts in column positions to effectively ‘reverse’ the distortion and reconstruct the higher symmetry parent structure.
}
Because the Fourier transform amplitude is only altered at the superlattice points and the phase is left unchanged, other changes to the experimental STEM image will be minimal. 
Features such as interfaces, boundaries, defects, strains, and even other symmetry breaking distortions with different wavevectors will be preserved in the resulting reference image.

Following the workflow described in Ref. \cite{savitzky2017bending},  we demonstrate in Figure \ref{fig:F2} how this approach enables the measurement of picometer-scale periodic lattice displacements in the manganese oxide \SBMO. 
\SBMO crystallizes with a double perovskite structure where samarium and barium alternate on the \textit{A}-site along [001] \cite{troyanchukNewFamilyLnBaMn2O62002}. When cooled the material develops symmetry breaking charge order and associated atomic displacements \cite{arima2002prb}.
Figure \ref{fig:F2}a shows an atomic resolution HAADF-STEM image acquired on the [001] zone axis: the \textit{A}-site alternation is lost in projection, but the positions of the atomic columns are modulated by displacements (too small to see by eye) associated with charge order, which give rise to superlattice peaks in the image's Fourier transform (Fig. \ref{fig:F2}b). 
The magnitude of the Fourier transform is damped to the background level in circular areas around each superlattice peak while the phase is left unchanged (Fig. \ref{fig:F2}c).
Taking the real part of the inverse Fourier transform results in an image almost identical to the original, but with minute shifts in the positions of the atomic columns towards their undistorted, high-symmetry positions (Fig. \ref{fig:F2}d). 
Mapping out the change in the image contrast by taking the difference in the intensity of each pixel between the original image and the high-symmetry reference reveals a periodic pattern of dumbbell-like features centered on the atomic columns (Fig. \ref{fig:F2}e). 
These dumbbells arise from shifts of the atomic column positions smaller than the widths of the columns themselves, which result in reduced intensity on one side of the column and increased intensity on the other. 
These shifts are quantified by fitting 2D Gaussian functions to each atomic column in both the original and reference image to precisely determine their positions. 
The vectors from the position of each column in the reference to the corresponding position in the original image describe the atomic displacements associated with charge order in the material and can be mapped to visualize the structural changes in the material attributable to the broken symmetry (Fig. \ref{fig:F2}f).
\textcolor{black}{
Construction of the reference image leverages the clean separation of lattice and superlattice signals in Fourier space.
Here that construction is performed by damping the superlattice peaks;
}
\textcolor{black}{
similar analysis can also be performed via filtering in Fourier space to select only the superlattice peaks, rather than damping them. This will result in a similar signal to Figure \ref{fig:F2}e, which can be subtracted from the original image to arrive at a high symmetry reference image
}
\textcolor{black}{
as discussed in Supplementary Note IV \cite{supp}. 
}
\begin{figure}[ht]
    \includegraphics[clip=true,width=\columnwidth]{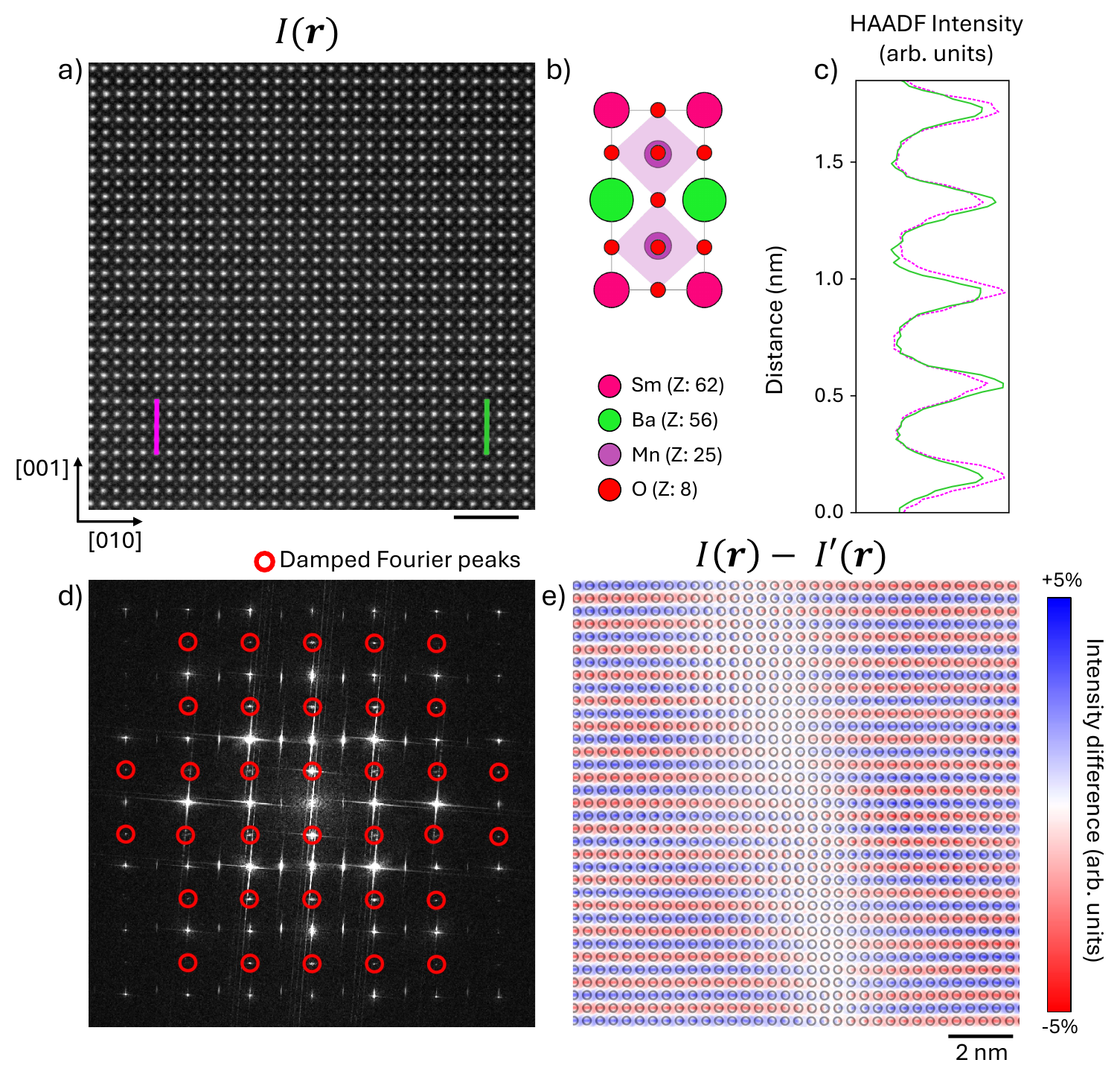}
    \caption{Visualizing chemical order with Fourier damping. 
    a) A cropped region of an atomic-resolution HAADF-STEM image of a \SBMO thin film on the [100]$_pc$ zone axis. 
    b) The \SBMO unit cell showing the alternating layers of samarium and barium occupation of the perovskite $A$-sites. 
    c) Profiles of image intensity from the left (pink) and right(green) sides of the image along lines marked on (a). 
    d) Fourier transform of (a), peaks corresponding to the alternating samarium and barium layers marked in red.
    e) Intensity differences between the original image (a) and a reference generated by damping the peaks marked in (d) to suppress the chemical order, showing alternating intensity peaked on the $A$-sites (indicated by circular markers) and an anti-phase boundary in the chemical order.
    }
     \label{fig:F3}
\end{figure}

\textcolor{black}{
The Fourier damping approach
}
has been extensively applied to mapping the atomic displacements associated with charge order \cite{savitzky2017bending,el2018nature,el2021charge,scheid2025direct,schnitzer2025atomic}, as well as antipolar displacements \cite{goodge2022disentangling,xuTwoDimensionalAntiferroelectricityNanostripeOrdered2020}, and even artifacts from channeling of the electron beam \cite{smeaton2023influence}, but it can also be readily generalized to characterizing other kinds of symmetry breaking that do not result in a clear displacement of atomic columns. 
Any periodic symmetry breaking with a non-zero wavevector (i.e. which generates superlattice peaks) that alters the image contrast can be characterized with this technique, meaning it can be applied to chemical and vacancy order, as well as distortions which may not be coherent along the beam direction \cite{el2020direct}.
Such distortions can be recognized by the absence of the characteristic dumbbell pattern associated with atomic column displacements seen in Figure \ref{fig:F2}e, and while similar image analysis can be applied it requires different interpretation and visualization than that shown in Figure \ref{fig:F2}.
Figure \ref{fig:F3} demonstrates these distinctions with an analysis of the chemical order also present in \SBMO. 
Viewed along the [100] zone axis, alternation of the \textit{A}-site cation between samarium ($Z = 62$) and barium ($Z = 56$) creates a layered structure (Fig. \ref{fig:F3}b), with a small but discernible modulation in the atomic number dependent contrast of HAADF-STEM imaging (Fig. \ref{fig:F3}a). 
This ordering can be identified by taking line profiles along the \textit{A}-site cation columns (Fig. \ref{fig:F3}c), but visualizing inhomogeneity in the chemical order requires mapping the associated intensity modulations across full STEM images.
Extensive analysis of local intensity profiles would be tedious at best, and at worst computationally intractable to do over large areas with sufficient real space sensitivity.

The alternation of samarium and barium columns doubles the size of the unit cell along the $c$-axis, generating superlattice peaks in reciprocal space with a wavevector of $\frac{1}{2}c^*$ (Fig. \ref{fig:F3}d). 
As in the case of the periodic lattice displacements induced by charge order, masking these peaks results in a reference image where the symmetry breaking distortion - in this case, the modulation in $A$-site contrast - is suppressed. 
Figure \ref{fig:F3}e shows the difference between the original STEM image and this high symmetry reference, which reveals the layers of alternating contrast in the \textit{A}-site planes, where blue indicates higher intensity (i.e., Sm) and red a lower intensity (i.e., Ba), but lacks the dumbbell features associated with atomic column displacements seen previously (Fig. \ref{fig:F2}e).
Mapping out the chemical order in this way also reveals an anti-phase boundary in the order, clearly visible extending vertically in the difference map. 
The presence of this defect can be confirmed by comparing intensity profiles measured on either side of the STEM image which show opposite intensity modulations on the $A$-site columns (Fig. \ref{fig:F3}c) \textcolor{black}{or comparing fit intensities of each atomic column in the image (Suppl. Fig. 3) \cite{supp}}.
By mapping distortions continuously in real space with Fourier damping, however, defects can be identified without prior knowledge of their presence and inhomogeneity over large length scales can be systematically measured. 

\textcolor{black}{
For reliable analysis, understanding the effects of parameter choices on the image processing is critical.
Here, a set of superlattice peaks are identified and the amplitude of circular regions of the Fourier transform around each peak are set uniformly to the local background level. 
Because the Fourier transform background level varies at different wavevectors, we first fit each peak with a 2D Gaussian function, and set the circular damping area to the fitted background. 
This leaves two key remaining parameters for the analysis: 
}
(1) the number of superlattice peaks damped, and (2) the size of the region damped around each peak. 
Varying the number of peaks damped controls what information is mapped back to real space and can affect the accuracy of measured distortions (Suppl. Fig. 2) \cite{supp}. 
The presence of inhomogeneity, including defects like the anti-phase boundary shown in Figure \ref{fig:F3}, will broaden the superlattice peaks associated with symmetry-breaking distortions. 
The information in the Fourier transform which encodes the morphology of the order will extend away from the reciprocal lattice points into their peak tails, meaning that the size of the damped reciprocal space regions  is also an important parameter to consider when conducting this analysis. 

\begin{figure*}[ht]
    \includegraphics[clip=true,width=\textwidth]{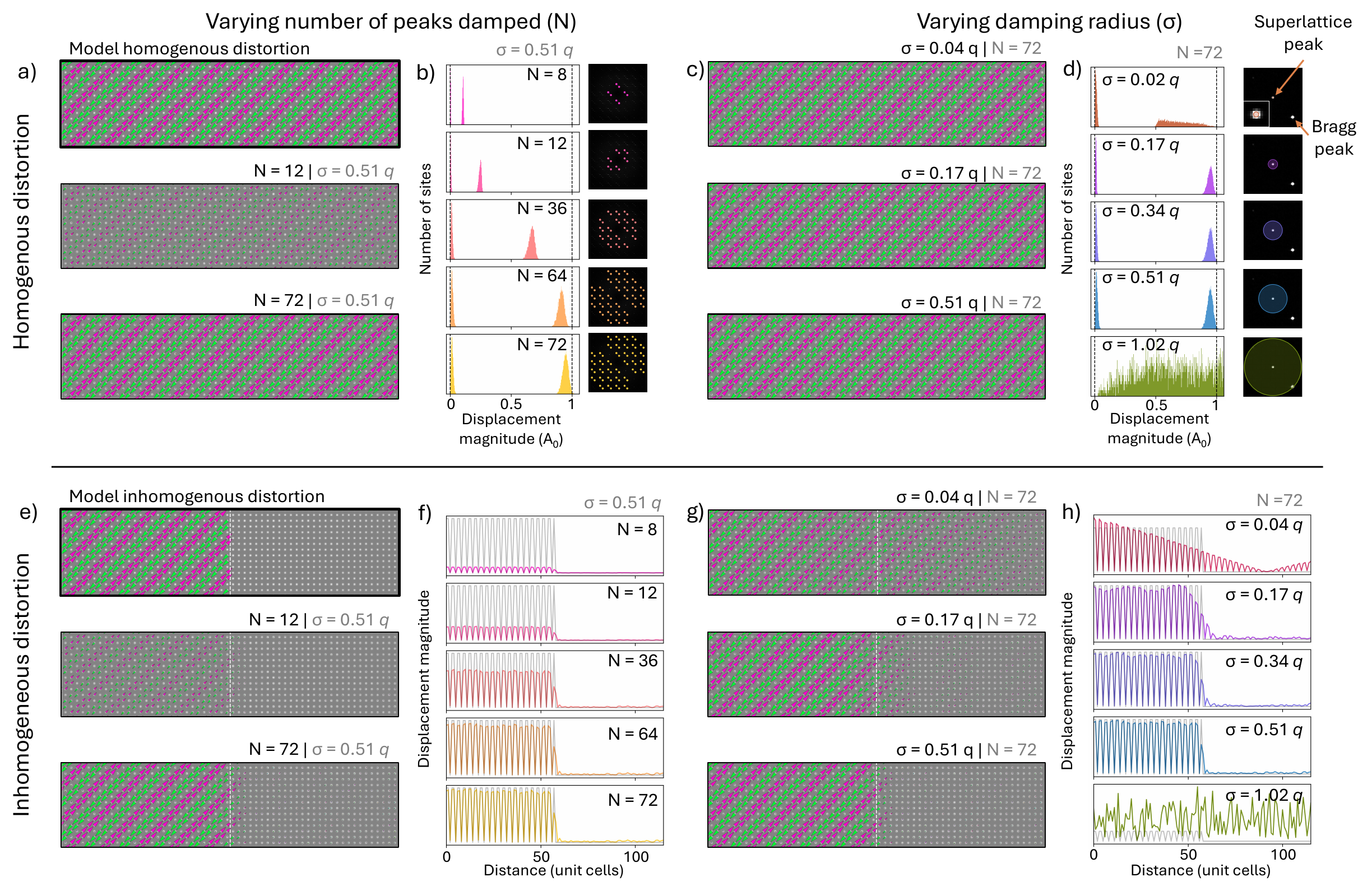}
    \caption{Validating Fourier damping on simulated datasets. 
    a) A model structure uniformly distorted with periodic lattice displacements (top panel), and maps of displacements measured by damping only superlattice peaks around the first and second order Bragg peaks (middle panel) and all superlattice peaks visible (bottom panel). 
    b) Histograms (left) of displacement magnitudes showing increasing measured displacements as the number of peaks damped is increased. Black dashed lines at 0 and A$_0$ indicate the true displacement magnitudes in the modeled structure. Fourier transforms (right) show the peaks damped in each case.
    c) Maps of measured displacements for small (top panel), intermediate (middle panel), and large (bottom panel) radii damped around each peak, while the number of peaks masked is held constant. 
    d) Histograms (left) showing the measured displacement magnitudes for various radii damped around the superlattice peaks. Fourier transforms cropped to center on a superlattice peak show the radii used in each case.
    e) A model structure with a distorted domain (left) and undistorted domain (right) separated by an atomically sharp interface, and maps of displacements measured damping only low order peaks (middle) and all peaks (bottom).
    f) Profiles of displacement magnitudes across the interface. Colored lines show magnitudes measured with various numbers of peaks damped, gray lines show the true modeled displacement magnitudes. 
    g) Maps of measured displacements for small (top), intermediate (middle), and large (bottom) radii damped around each peak showing increasing interface sharpness.
    h) Profiles of displacement magnitudes across the interface for various damping radii (colored lines) and the true modeled displacements (gray). 
    }
    \label{fig:F4}
\end{figure*}
%TODO: add N=2?
The quantitative dependence of Fourier damping measurements on these parameters -- size and number of damping masks -- is assessed on simulated STEM images with known atomic displacements. 
Varying both the number of peaks ($N$) and the radius of the circular region of reciprocal space about each peak ($\sigma$) which are damped, displacements are first measured on a test image of a structure modulated by uniform sinusoidal atomic displacements similar to those shown in Figure \ref{fig:F2}. 
Damping fewer superlattice peaks reduces the measured amplitude of the atomic displacements, as can be seen in Figure \ref{fig:F4}a. 
As the number of damped peaks increases, the measured displacement amplitude approaches – but remains systematically smaller than – that of the true displacements in the structure (Fig. \ref{fig:F4}b). 
Thus, damping a large set of peaks including those at large angles and higher order satellites - in practice, as many superlattice peaks as can be identified in the Fourier transform - is critical to estimate the amplitude of atomic displacements, and such measurements should be interpreted as a lower bound on the true displacement.

On the other hand, for a uniform modulation changing the damping radius has little effect, unless it is made so small that most of the peak magnitude is not masked at all, or so large that other peaks, such as the neighboring Bragg peaks, are also damped. 
In either of these extremes the resulting displacements are entirely incorrect, but for a large range of intermediate values the displacements are accurately measured (Fig. \ref{fig:F4}c,d). 
When the distortion incorporates significant inhomogeneity, however, the damping radius becomes more significant. 
The top panel of Figure \ref{fig:F4}e shows a model structure with an atomically sharp interface between an area with periodic lattice displacements and an area without. 
As in the case of a uniform modulation, damping more or fewer superlattice peaks alters the amplitude of the measured displacements, but has little effect on the sharpness of the interface (Fig. \ref{fig:F4}e,f). 
On the other hand, adjusting the damping radius has a strong effect on the sharpness with which the interface is resolved. 
As can be seen in Figure \ref{fig:F4}g, when only a very small area around the peak center is damped, the interface is not resolved at all.
Rather, distorted and undistorted domains are only hinted by a gradual ramp in the displacement amplitude. 
As the damping radius is increased, the spatial resolution is  enhanced and the measurement quickly approaches the model structure. 
Over a broad range of radii ($\sigma=0.34q$ to $0.51q$) the measurement is sharp and accurate to the model distortion, indicating that even for \textcolor{black}{inhomogeneous} distortions a wide range of parameters will provide reliable measurement. 
Care must nonetheless still be taken that the radius is not made so large that other peaks (or even their tails) are included within it, as this introduces serious artifacts and makes the measurement uninterpretable. 
While for the simple model presented here this effect is not apparent until an extremely large radius impinges directly on the Bragg peak ($\sigma=1.02$), when applying the technique attention should be paid to what features are damped and various damping radii should be tested for consistency.
\textcolor{black}{
Similar analysis of parameter choices on experimental datasets is shown in Supplemental Figure 8 \cite{supp}.
}
Full two-dimensional sweeps through both parameters are included in Supplementary Note III \cite{supp}.

Line profiles of the displacement magnitude across the interfaces make the trends even more clear - at low damping radii, the displacement profiles resemble a Fourier series with very few terms.
As more pixels of the Fourier transform are damped, the series converges towards accurately representing the sharp interface (Fig. \ref{fig:F4}f,h). 
This result can be understood by considering the delocalization of information in reciprocal space in the presence of inhomogeneity: a real-valued plane wave involves only a single spatial frequency or one Fourier coefficient, and in reciprocal space it is represented by a conjugate pair of delta functions. 
When the wavelength and amplitude of the wave are allowed to spatially vary, however, the wave will extend over a range of frequencies in reciprocal space and require representation with multiple Fourier coefficients. 
As the sharpness of these features increases, more coefficients are needed for an accurate representation. 
Consequently, the damping radius is inversely related to coarsening in real space, so large damping radii must be used for accurately measuring spatial variations in inhomogeneous  distortions, 
\textcolor{black}{
just as larger Fourier filters result in a smaller coarsening length (Supplementary Note I) \cite{supp}. 
}
This highlights a constraint on the technique: large-wavelength or low-frequency distortions will have small wavevectors and little separation between superlattice and Bragg peaks in the Fourier transform, limiting the achievable spatial resolution for such measurements \cite{savitzky2017bending}.

The Fourier damping approach presented here enables robust measurement of symmetry breaking distortions in atomic-resolution images through the generation of self-consistent high-symmetry reference images. 
By taking advantage of the localization of these distortions in reciprocal space, a wide variety of phenomena including periodic lattice displacements, chemical order, vacancy order, and more can be analyzed, and the measurements directly represent physically meaningful changes in symmetry. 
Coexisting distortions associated with distinct wavevectors can be decoupled, as demonstrated for instance in Ref. \cite{savitzky2017bending}.
Clear heuristics for parameter selection are established and benchmarking on model datasets with known distortions demonstrates the high accuracy and precision achievable with this technique.
\begin{figure*}[ht]
    \includegraphics[clip=true,width=\textwidth]{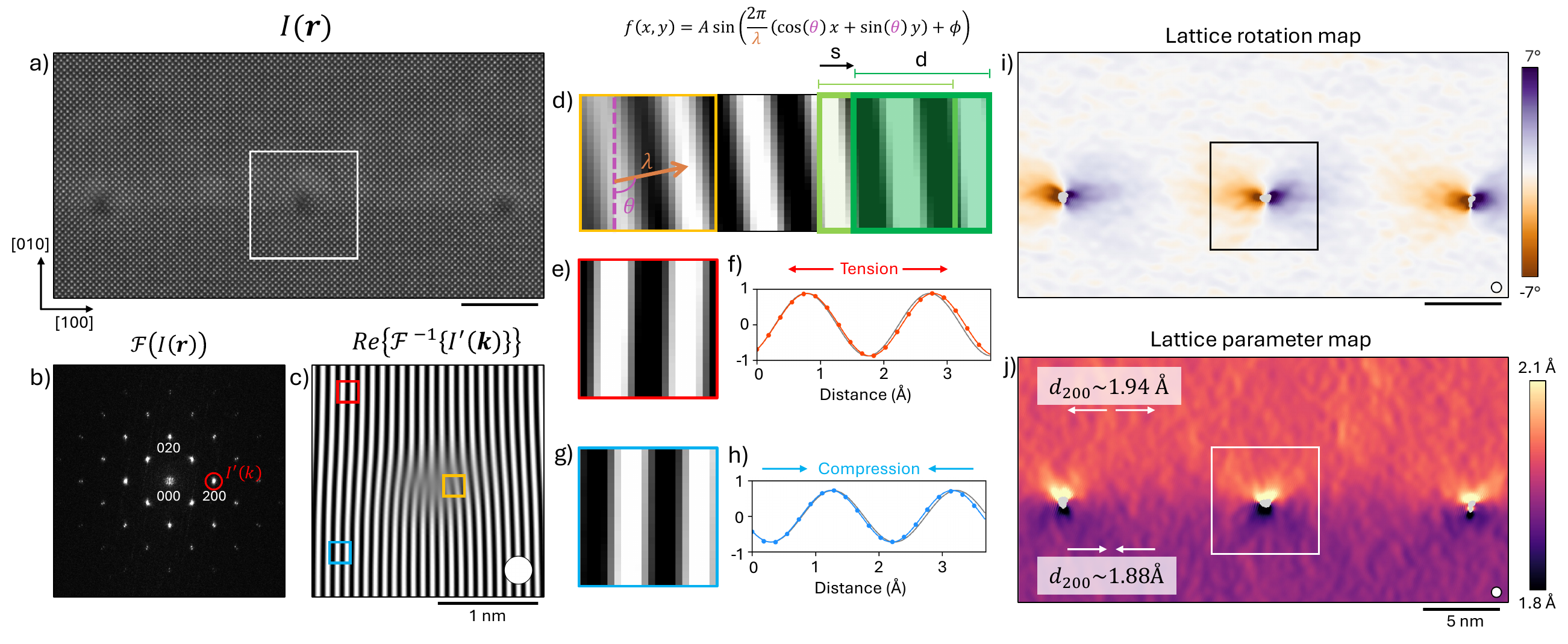}
    \caption{Measuring lattice parameters with local wave fitting. 
    a) An atomic resolution HAADF-STEM image of the interface between a \SLAO substrate and a relaxed \SRO thin film, with misfit dislocations visible along the interface.
    b) Fourier transform of the image, with the lowest order in-plane Bragg peak (200) marked in red. 
    c) The real component of the Fourier filtered signal from the peak marked in (b) over the area marked by the white square in (a), showing sinusoidal fringes corresponding to the (200) lattice planes with additional half planes at the site of the misfit dislocation.
    d) Patches of the Fourier filtered signal. Annotations mark the wavelength (orange arrow) and orientation (purple markings) of the fringes visible in a patch. Patches can overlap with a spacing \textit{s} smaller than their dimension \textit{d}. 
    e) A patch of the Fourier filtered signal from the film (marked in red on (c)), with a larger inter-planar spacing (tensile strain) relative to average across the full image.
    f) Projection of the patch shown in (e) to one dimension (red dots), fit to a sine wave (red line) and the average periodicity of the structure (grey line).
    g) A patch of the Fourier filtered signal from the substrate (marked in blue on (c)), with a smaller inter-planar spacing (compressive strain) relative to average across the full image.
    h) Projection of the patch shown in (g) to one dimension (blue dots), fit to a sine wave (blue line) and average periodicity of the structure (grey line).
    i) Map of the lattice rotation across the same field of view shown in (a).
    j) Map of the in-plane lattice parameter ($2 \times d_{200}$) across the same field of view shown in (a).
    Coarsening lengths associated with Fourier filtering and local wave fitting are marked by the diameters of white circles, as described in Supplementary Notes 1 and 2 \cite{supp}.
    }
     \label{fig:F5}
\end{figure*}

\section{\label{sec:wavefit}Local wave fitting for lattice parameter measurement}
While damping information in Fourier space is useful for revealing superlattices associated with discrete broken symmetries, in many systems, continuous deformations which locally alter lattice parameters are key to understanding functional properties.
The combination of high resolution and high precision make STEM a powerful technique for measuring such spatial variations in inter-planar spacings and angles. 
Often parameterized as strains - relative measures of deformation compared to a reference region or nominal value - an understanding of these variations on the nanometer- to micron-scale is frequently a primary goal of STEM analysis. 
Applications include identifying defects, checking for relaxation across interfaces, differentiating domains associated with twinning or ferroelasticity, identifying particular components of nanostructures, and device metrology \cite{chung2007measurement,zhu2013interface,hytch2006stress,gao2014ferroelastic,haussler2007quantitative,chung2010practical}. 

In an atomic resolution STEM image, local strain can be measured simply by calculating the distances between neighboring atomic columns and comparing to a known value or reference region \cite{jia2007unit}. Mapping strain across a full image in this way, however, depends on precise identification of the positions of all of the columns in the image,  involving slow, resource intensive fitting of each column, such that success depends strongly on the resolution and signal-to-noise ratio of the image.
More commonly, strains are measured using GPA \cite{hytchGeometricPhaseAnalysis1997, hytch2003measurement} or phase-lock in approaches \cite{mesarosTopologicalDefectsCoupling2011,lawlerIntraunitcellElectronicNematicity2010,goodge2022disentangling}, which Fourier filter a single peak to select a periodicity of interest, calculate a geometric phase that captures the deviations from this periodicity, and measure strains from its gradient.

By eliminating the need for precisely fitting the positions of each atomic column, these approaches allow analysis to be easily scaled to larger areas, but also have their own limitations. 
One is that the quantification is relative to a reference periodicity, complicating absolute measurement of inter-planar spacings. 
Another is the sensitivity of these measurements to noise: if the periodicity being studied has a low signal-to-noise ratio (SNR) - for instance because its amplitude is suppressed from disorder - artifacts which resemble phase slips and dislocations but have no physical origin will appear due to the nature of the image processing performed \cite{schnitzer2025atomic}. 
Artifacts can also emerge from changing contrast of atomic columns, for instance due to different compositions across heterostructure interfaces \cite{petersArtefactsGeometricPhase2015}.
These issues are compounded by the lack of a means to readily measure the uncertainty or error associated with the measurement of the phase, meaning processing artifacts can easily be mistaken for real sample features.
On the other hand, with recent 4D-STEM detector development, strain is now commonly measured with scanning nanodiffraction, which can experimentally be very convenient and requires no atomic resolution imaging at all \cite{ozdol2015strain}. Still, the resolution of this technique is limited by the requirement of a small convergence angle to avoid overlap of diffraction discs, which can be  challenging to optimize especially in materials with large unit cells or highly localized strain variations. As the convergence angle must be selected prior to the measurement, fine features and the presence of defects which are clear in imaging may be missed.

Here, we present a technique for quantifying lattice spacings and angles from atomic resolution images by locally fitting waves in real space to a Fourier filtered signal, developed for analyzing challenging nanostructured materials \cite{smeaton2021mapping}.
While the Fourier processing is similar to GPA, by fitting in real space this approach offers distinct advantages. 
For instance, absolute inter-planar spacings are measured rather than relative strains, eliminating the need for a pristine reference area for use as a reference. Additionally, the fitting variance can be used as a heuristic for the uncertainty associated with the measurement, providing the error bars missing from techniques like GPA as well as a means of identifying and excluding artifacts and unreliable measurements.

Figure \ref{fig:F5} shows the steps involved in the local wave fitting algorithm, using as an example an image of a partially relaxed thin film of \SRO grown on a \SLAO substrate with an array of  misfit dislocations along the film-substrate interface. 
Fourier filtering the in-plane (200) peak by weighting the Fourier transform of the STEM image with a narrow Gaussian distribution centered on the peak of interest, such that only information about that periodicity is retained, and taking the real part of the inverse transform back to real space results in sinusoidal fringes peaked on each (200) plane of the lattice. 
As detailed in Supplementary Note I \cite{supp}, the spatial resolution or coarsening length associated with this Fourier filtering is inversely proportional to the width of the Gaussian distribution used.
Increasing the filter size may also incorporate more noise, reducing the SNR and potentially introducing artifacts. 
Thus, an appropriately sized filter - wide enough that a sufficient spatial resolution is achieved to map the features of interest, but not so wide that more than one peak or excessive noise is included - is critical for meaningful analysis.
As can be seen in Figure \ref{fig:F5}c which shows a portion of this fringe image around one of the misfit dislocations, variations in the periodicity of the wave reflect the different lattice spacings of the film and substrate, while the misfit dislocation is marked by additional half-planes and a bending of the surrounding fringes, matching the behavior of the atomic planes in the original image.

The fringe image is then divided into small patches, typically around twice the fringe wavelength, so that variations in the wavelength and angle of the fringes are apparent. 
To avoid degradation in resolution, the sampling of the patches is made at a greater frequency than the patch size, such that they overlap (Fig. \ref{fig:F5}d). 
The amplitude, wavelength, and orientation of the planes in each patch are found by fitting with 2D sine waves, allowing the spatial variations in each of these parameters to be mapped.
For instance, patches from the film have an increased wavelength compared to the average across the full field of view, which can be seen clearly when projected to one dimension (Fig. \ref{fig:F5}e,f), while fringes in the substrate are relatively compressed, reflecting its smaller in-plane lattice parameter (Fig. \ref{fig:F5}g,h).

By fitting these waves in patches across the whole image, the spacing and orientation of the (200) planes can be mapped out from the fit parameters  (Fig. \ref{fig:F5}i,j). 
The inter-planar spacing is directly measured in real space units, an important consideration when an understanding of absolute lattice parameters rather than relative strains is desired. 
In this case, the independent measurement of the film and substrate lattice parameters shows the expected uniform relaxation of the film suggested by the presence of the regularly spaced misfit dislocations, and an interesting visualization of the deformation surrounding the defects. By repeating the analysis with other peaks, other periodicities such as the out-of-plane spacing can be measured in the same way.

\begin{figure*}[ht]
    \includegraphics[clip=true,width=\textwidth]{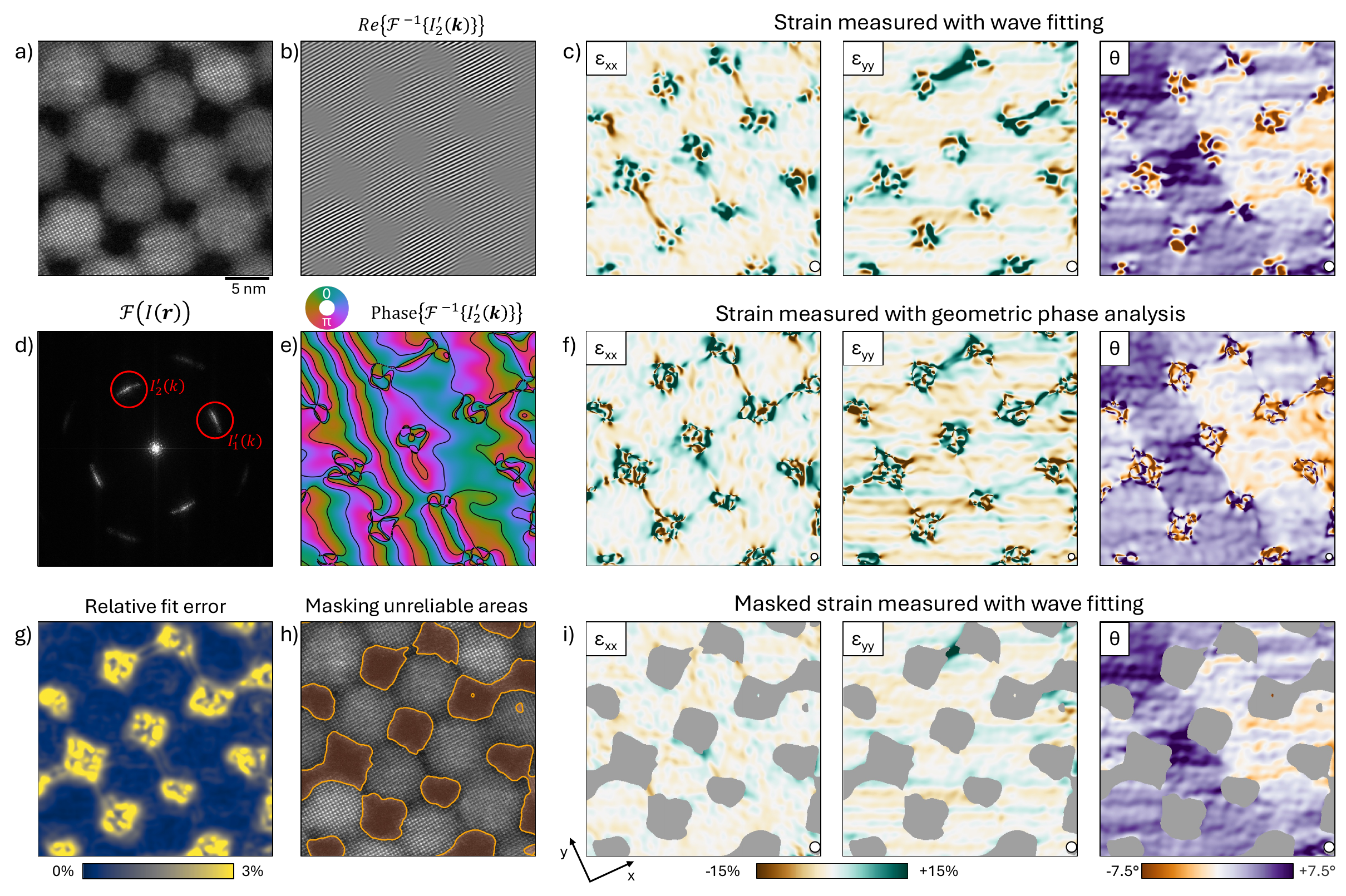}
    \caption{Local wave fitting application and comparison with geometric phase analysis.
    a) A HAADF-STEM image of a superlattice of PbSe quantum dots. 
    b) Fringes mapped from real part of Fourier transform of the $I_2$ peak marked in (d).
    c) Longitudinal strains (left and center) measured along two principal axes from wave fit d-spacings relative to the average spacings across the field of view,  and the local lattice rotation (right).
    d) Fourier transform of the STEM image, showing azimuthally elongated Bragg peaks due to orientational disorder in the superlattice. Superlattice peaks can be seen very close to the central peak but are not used in this analysis.
    e) Phase associated with the Bragg $I_2$ peak marked in (d) extracted with geometric phase analysis. Large phase variations arise from real strains in the system as well as from artifacts in areas without crystalline structure.
    f) Strains mapped with geometric phase analysis, showing strong similarity to the wave fit result.
    g) Relative errors of $\epsilon_{xx}$ and $\epsilon_{yy}$ summed in quadrature, which peaks in gaps between quantum dots without crystalline order.
    h) Mask created by thresholding the fit error which segments areas with low fit certainty.
    i) Strains measured with wave fitting and masked to include only areas with low uncertainty.
    \textcolor{black}{
    Coarsening lengths associated with the geometric phase analysis (3.6 Å) and local wave fitting (5.7 Å)  are marked by the diameters of white circles in the lower right of each panel, as discussed in Supplementary Notes 1 and 2 \cite{supp}.
    }
    }
     \label{fig:F6}
\end{figure*}

In cases where specifically a strain measurement is desired, the inter-planar spacings can be converted to relative strains simply by measuring the deviation in the measured wavelength relative to a chosen reference. To validate the approach, strain is measured in a superlattice of interconnected quantum dots \cite{mccrayOrientationalDisorderEpitaxially2019, smeaton2021mapping} with both the conventional GPA approach and the local wave fitting algorithm presented here \textcolor{black}{using identical 2D Gaussian filters}  (Fig. \ref{fig:F6}a).
The strains measured with each technique are very comparable, revealing in both cases the variations in particle orientation and localized strains present at the crystalline bridges between some dots (Fig. \ref{fig:F6}c,f).

This similarity is indeed unsurprising, as both measurements at a basic level are mapping phase variations in the same Fourier filtered signal. The main distinguishing feature of the different approaches is thus their  interpretability. 
A key advantage of the local wave fitting approach and motivation for its use is the extraction of not only spacing and rotation of lattice planes, but also variances of the associated fits. 
These variances give valuable information about the error associated with lattice parameter or strain measurements, which is missing from conventional approaches like GPA and phase lock-in. 
This is particularly important when studying inhomogeneous systems, such as nanostructures with distinct contrast in different components, twin domains with peaks at distinct positions in reciprocal space, and any images with significant local disorder or otherwise poor signal-to-noise ratios - all of which can give rise to artifacts in the form of non-physical phase variations. 
These uncertainties can be used to set independent, quantitative error bounds on all measured strains and parameters, allowing significant and insignificant variations to be differentiated and comparisons with other characterization techniques such as electron or X-ray diffraction to be better understood. 
For instance, the uncertainties can be used to understand the error in the strain measured over continuous and non-continuous necks between a pairs of quantum dots, shown in Supplementary Note V \cite{supp}.

For another example, GPA and local wave fitting both measure large apparent strains and discontinuities  in the gaps between particles where there is no lattice present at all.
These large ``strains'' and apparent defects emerge because when the amplitude of the Fourier filtered signal is low, the image's phase information is dominated by noise, rather than real sample features. 
This noise causes the phase to vary rapidly, as can be seen in the image of the geometric phase extracted with GPA (Fig. \ref{fig:F6}e).
Because these rapid variations can look quite similar to large strains associated with dislocations and out-of-phase boundaries, hasty analysis may confuse them for real features.
The variance of the fits performed with wave fitting allow these unreliable areas to easily be identified and masked (Fig. \ref{fig:F6}h,i) in order to visualize only the meaningful strain information extracted from the image (Fig. \ref{fig:F6}j).
While a similar mask could be created in this case by using the amplitude of the Fourier filtered signal, the rich, quantitative information provided by the fit uncertainty offers widely useful information for the interpretation of measured strains, a key advantage that other similar methods lack.

\section{\label{sec:Discussion}Discussion}
This work has presented approaches for reliable quantification of STEM images using a synergy of real and reciprocal space information to address two key structural characterization problems – the measurement of translational symmetry breaking distortions and local variations in lattice parameters. 
Taking advantage of the localization of information associated with symmetry breaking in reciprocal space, the Fourier damping algorithm facilitates the high-precision measurement of periodic lattice displacements, as well as a variety of other distortions including chemical and vacancy orderings. 
By enabling direct comparison between distorted and parent structures, the technique facilitates physically meaningful analysis. This ability to separate distortions of interest from lattice disorder and other coincident orderings  helps to disentangle the structural complexities found in many functional materials.
Similarly, the local wave fitting approach for lattice parameter quantification offers a convenient means for absolute measurement of inter-planar spacings and angles, and comparable strain measurements to common approaches such as GPA, while providing an enhanced understanding of the uncertainty associated with these measurements.
These algorithms, along with a suite of other atomic resolution image processing routines including phase-lock in analysis, calculation of correlation functions, and other local crystallographic techniques have been implemented and made available to the community in the open-source \texttt{kemstem} Python package.

With the growing adoption of atomic resolution STEM characterization in the materials development cycle, the development of robust analysis techniques capable of reliably extracting the maximum available structural information from the resulting imaging data is essential. 
While advances in computing hardware and statistical analysis approaches enable the study of increasingly large and complex datasets, care must nonetheless be taken to ensure that the resulting insights are physically meaningful. 
The techniques presented here show how applying physically informed analysis to harness both the real and reciprocal space information encoded in atomic resolution STEM images can enable robust quantification of features of interest while transparently revealing the associated sources of error and levels of uncertainty.
Continuing development in these areas will further improve the reliability and scalability of atomic resolution image analysis to introduce greater automation both for post-hoc analysis and real-time on-line processing, as well as adapting approaches to take full advantage of hardware developments in general purpose GPU computing.

\begin{acknowledgments}
The techniques presented in this work and implemented in the \texttt{kemstem} package were developed in the Kourkoutis Electron Microscopy (KEM) group at Cornell University under the leadership of the late Professor Lena F. Kourkoutis. 
The authors acknowledge her significant direction, guidance and support, without which this work would not have been possible.
The authors also acknowledge C.T. Parzyck, K.M. Shen, S. Yamada, T. Arima, Y.A. Birkhölzer, N. Wadehra, D.G. Schlom, D. Balazs, and T. Hanrath  for providing the samples for STEM imaging, and S.H. Sung for helpful discussion.
This work made use of the electron microscopy facility of the Platform for the Accelerated Realization, Analysis, and Discovery of Interface Materials (PARADIM), which is supported by the National Science Foundation under Cooperative Agreement No. DMR-2039380 and the Cornell Center for Materials Research shared instrumentation facility. The FEI Titan Themis 300 was acquired through NSF-MRI-1429155, with additional support from Cornell University, the Weill Institute and the Kavli Institute at Cornell.
\end{acknowledgments}

\section{Data Availability}
The datasets analyzed in this study are available from the Platform for the Accelerated Realization, Analysis, and Discovery of Interface Materials (PARADIM) database at \url{https://doi.org/10.34863/mjbn-ym63} \cite{data}.

\section{Code Availability}
The \texttt{kemstem} Python package is available at \url{https://github.com/noahschnitzer/kemstem}.

\bibliography{refs}

\end{document}

\title{Supplementary Information: \mytitle}

\author{Noah Schnitzer}
\email{nis29@cornell.edu}
\affiliation{\cmse}
\affiliation{\kav}

\author{Lopa Bhatt}
\affiliation{\caep}

\author{Ismail El Baggari}
\affiliation{\rowland}

\author{Robert Hovden}
\affiliation{\ummse}

\author{Benjamin H. Savitzky}
\email{ben.savitzky@gmail.com}
\affiliation{\caep}

\author{Michelle A. Smeaton}
\affiliation{\nrel}

\author{Berit H. Goodge}
\email{Berit.Goodge@cpfs.mpg.de}
\affiliation{\CPfS}

\date{\today}
\renewcommand{\figurename}{Supp. Fig.}

\maketitle

\section{Coarsening length associated with Fourier filtering}
When Fourier filtering an image ($I(\br)$), the Fourier transform of the image ($I(\bk) = \FT\{I(\br)\}$) is weighted with a filtering function ($W(\bk)$) centered on a peak of interest, and the inverse Fourier transform is taken to return to real space:
$$I_{\text{filt}}(\br) = \FT^{-1}\{ W(\bk) \cdot I(\bk)  \}$$
By the Fourier convolution theorem, the Fourier transform of this filtering function will be convolved with the image signal:
$$I_{\text{filt}}(\br) = \FT^{-1}\{W(\bk)\} \ast \FT^{-1}\{I(\bk)\} $$
meaning the filtering function will apply a real space coarsening to the filtered signal, with a profile set by the Fourier transform of the filtering function. 
This means that a sharp filter like a circular mask will introduce ringing features, while using a smooth Gaussian distribution is convenient as its Fourier transform is also a smooth Gaussian.
For a discretely sampled signal with dimension $D$, taking the Fourier transform of a Gaussian distribution $\mathcal{N}$ with standard deviation $\sigma_0$ will result in a Gaussian with $\sigma'=\frac{D}{2\pi\sigma_0}$. Thus, for a filtering function $W(\bk) = \mathcal{N}(\sigma_0)$, neglecting pre-factors the real space coarsening will be described by:
$$\FT^{-1}\{(W(\bk)\} \propto \FT^{-1}\{\mathcal{N}(\sigma_0)\} \propto \mathcal{N}(\frac{D}{2\pi\sigma_0})$$

This is shown graphically in Supplementary Figure \ref{sfig:coarsening}a. The standard deviation of the Gaussian profile, however, is not necessarily the most meaningful quantity for describing the information transfer through the filter. The full-width at half-maximum (FWHM) measures the width of the distribution at a broader point incorporating more of the total signal, and we choose this point to describe the coarsening length associated with Fourier filtering, as shown in Supplementary Figure \ref{sfig:coarsening}b. For a Gaussian filter with width $\sigma_0$ and a dimension $D$, the real space coarsening length parameterized as the FWHM of its Fourier transform is given by:
$$d_\circ=\frac{D}{2\pi \sigma_0}2\sqrt{\ln{4}}$$
This length is marked by the diameter of a circle in the lower-right corner of Fourier filtered images.

\section{Coarsening length associated with local wave fitting}
The coarsening length associated with local wave fitting is set by that of the initial Fourier filtering as well as additional loss of resolution in the subsequent processing steps. 
As discussed above, for image dimension $D$ and filter standard deviation $\sigma_0$ the Fourier filtered signal is taken to have a coarsening length: $$d_\circ=\frac{D}{2\pi \sigma_0}2\sqrt{\ln{4}}$$
This signal is subdivided into patches of dimension $d$, sampled with a step size $s$. Typically, $s < d \simeq d_\circ$, where the greater and more limiting of $d$ and $d_\circ$ will depend on situational parameter choices.
We take the larger as the coarsening length associated with local wave fitting, and mark it with the diameter of the circle shown in the lower-right corner of wave fit images. 
To aid comparison with the original image and avoid emphasizing local variations from fit errors, results are generally interpolated to match the original image dimensions, and smoothed with a filter narrower than this coarsening length.

\section{Parameter dependence of Fourier damping}
Two key parameters for Fourier damping are the number of peaks ($N$) and the radius of the circular region of reciprocal space about each peak ($\sigma$) damped, both of which control what reciprocal space information is incorporated into the analysis.
This is demonstrated in Supplemental Figure \ref{sfig:num_peaks}, which shows how varying the number of peaks damped alters the mapping of cation order shown in Figure 3.
Full simultaneous sweeps through the number of peaks damped and the damping radius corresponding to Figure 4 are shown for the model homogenous distortion (Suppl. Figs. \ref{sfig:sim_homog_maps}, \ref{sfig:sim_homog_hists}) and model distortion incorporating a sharp interface (Suppl. Figs. \ref{sfig:sim_inhomog_maps}, \ref{sfig:sim_inhomog_profiles}).

\section{Comparison of Fourier damping and filtering for reference image construction}
\textcolor{black}{
While Fourier damping and conventional Fourier filtering can both be applied to isolate information associated with translational symmetry breaking distortions and construct high symmetry reference images, their numerical details and physical intuitions differ somewhat. Numerically, damping provides an opportunity to more carefully handle the Fourier space background signal, mitigating a potential source of noise and artifacts. Geometrically, the damped transform represents precisely the Fourier decomposition of the original lattice image sans distortion components, i.e. the reference image, whereas Fourier filtering enables reconstruction of such an image slightly less directly taking advantage of the linearity of the Fourier transform. For comparison, the same dataset analyzed with Fourier damping in Figure 2 of the main text is processed with Fourier filtering in Supplementary Figure \ref{sfig:filter_compare}. 
}

\section{Visualizing uncertainty in measured parameters with local wave fitting}
A key advantage of the local wave fitting approach is its quantification of measurement uncertainty through the variance of the non-linear least squares fits performed on each image patch. In addition to masking out unreliable measurement areas, as shown in the main text, this understanding of spatial variations in measurement uncertainty can also provide error bars for the measured parameters, helping indicate whether variations are significant and measurements meaningful.

This is demonstrated in Supplemental Figure \ref{sfig:wf_errorbars} on the same dataset shown in Figure 6 of the main text,  visualizing the variance of the fit strains and quantum dot rotation along line profiles. The uncertainty measured across a neck between dots where the lattice is continuous (green) is smaller than that measured across a discontinuous neck (pink).

\bibliography{refs}

\begin{figure*}[ht]
    \includegraphics[clip=true,width=0.9\textwidth]{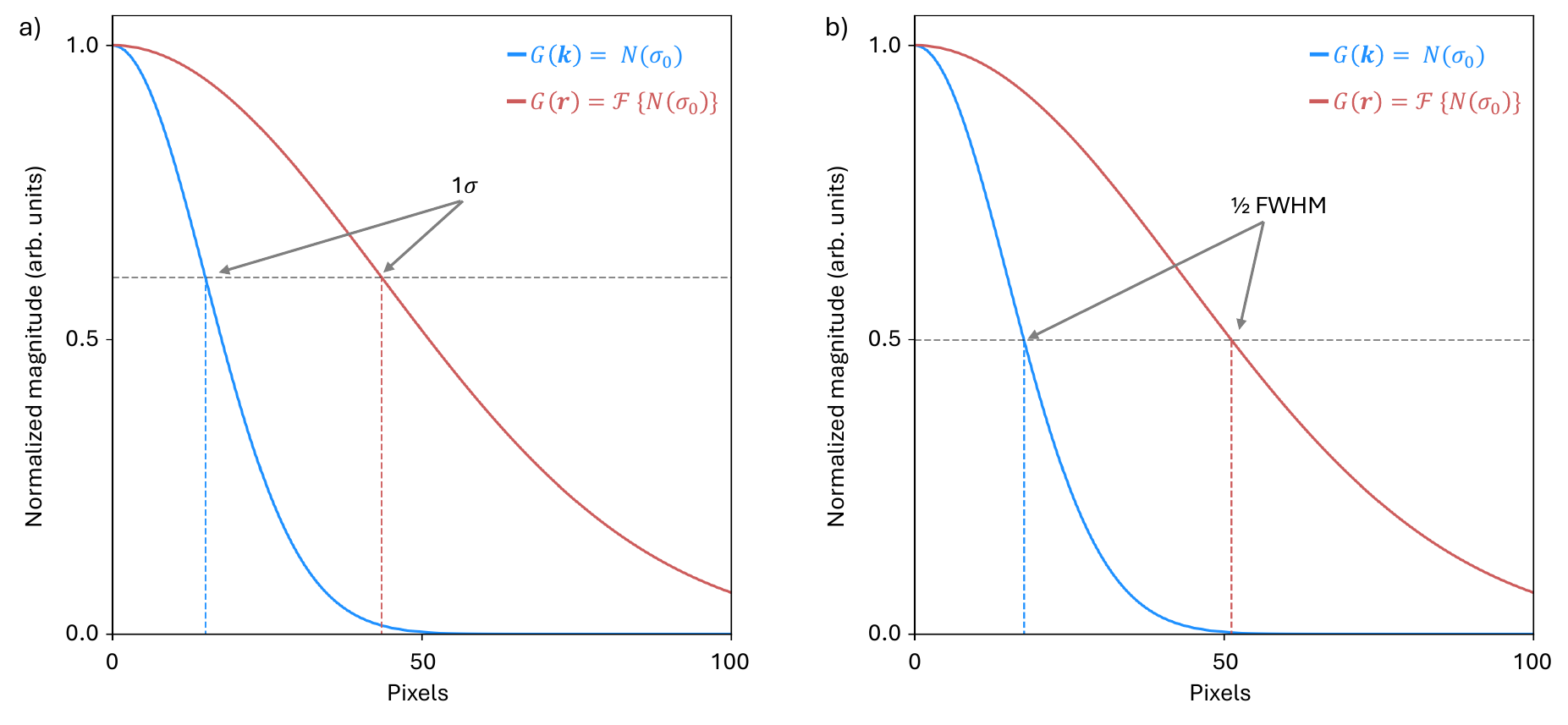}
    \caption{
    a) Standard deviation of a Gaussian distribution and its discrete Fourier transform. 
    b) Full-width at half-maximum of the same distribution and its Fourier transform.
    The x-axes are in pixels, for an image with dimension $a$ the real space pixel size will be $a/D$ and the reciprocal space pixel size will be $1/a$. Here, a signal with dimension $D=4096$ pixels and standard deviation $\sigma_0 = 15$ were used.
    }
    \label{sfig:coarsening}
\end{figure*}

\begin{figure*}[ht]
    \includegraphics[clip=true,width=0.65\textwidth]{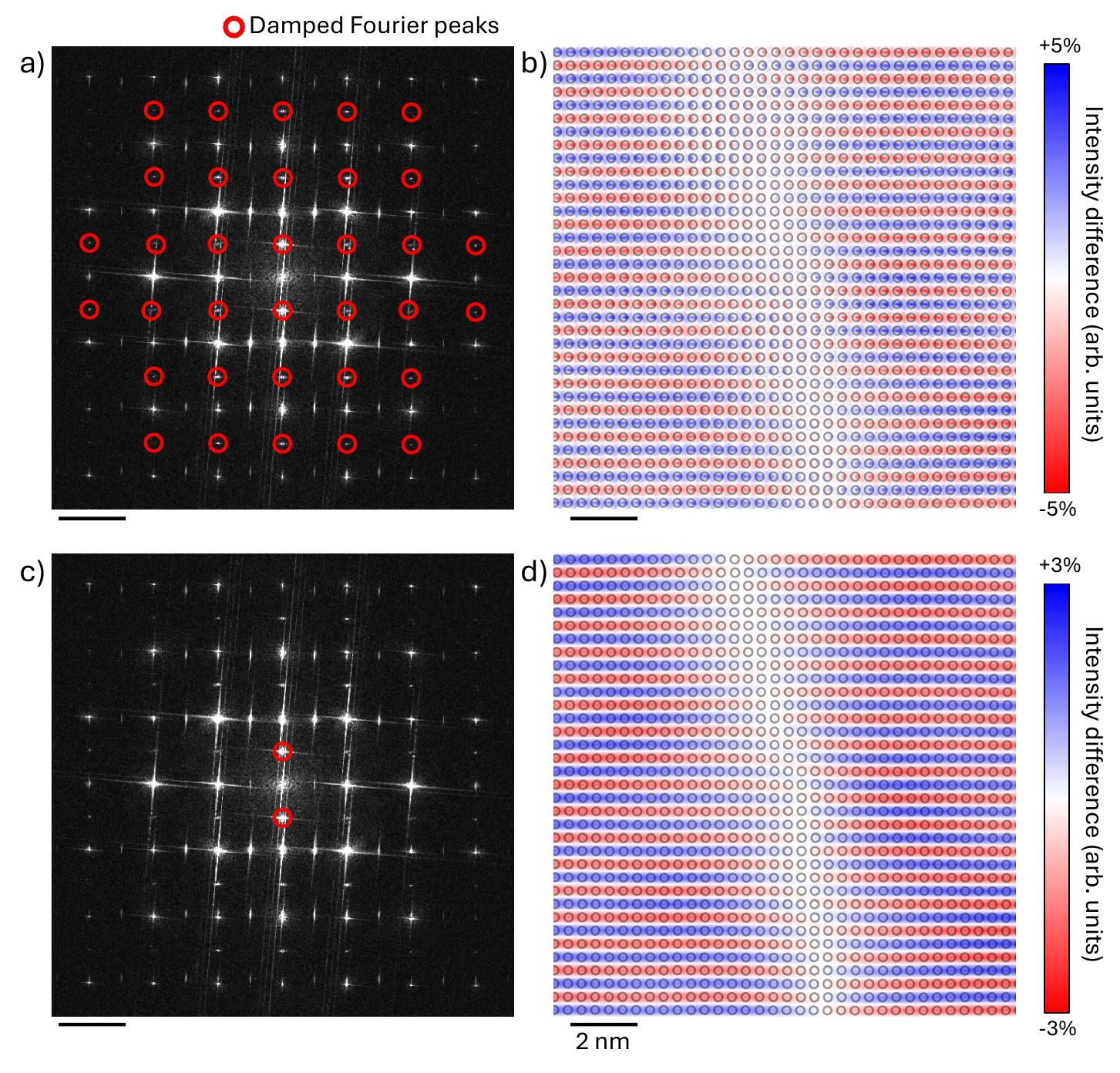}
    \caption{Visualizing chemical order damping various numbers of peaks.
    a) Fourier transform of the HAADF-STEM image shown in Figure 3a, with all low peaks corresponding to the alternating samarium and Barium layers marked in red.
    b) Intensity differences between the original image and a reference generated by damping the peaks marked in (a) to suppress the chemical order, showing alternating intensity peaked on the $A$-sites (indicated by circular markers) and an anti-phase boundary in the chemical order.
    c) Fourier transform of the HAADF-STEM image, with only the lowest order superlattice peak conjugate pair marked.
    d) Intensity differences from damping only the lowest order peaks marked in (c), resulting in a simple sinusoidal intensity variation peaked on the A-site layers, where the anti-phase boundary is still visible.
    }
    \label{sfig:num_peaks}
\end{figure*}

\begin{figure*}[ht]
    \includegraphics[clip=true,width=0.95\textwidth]{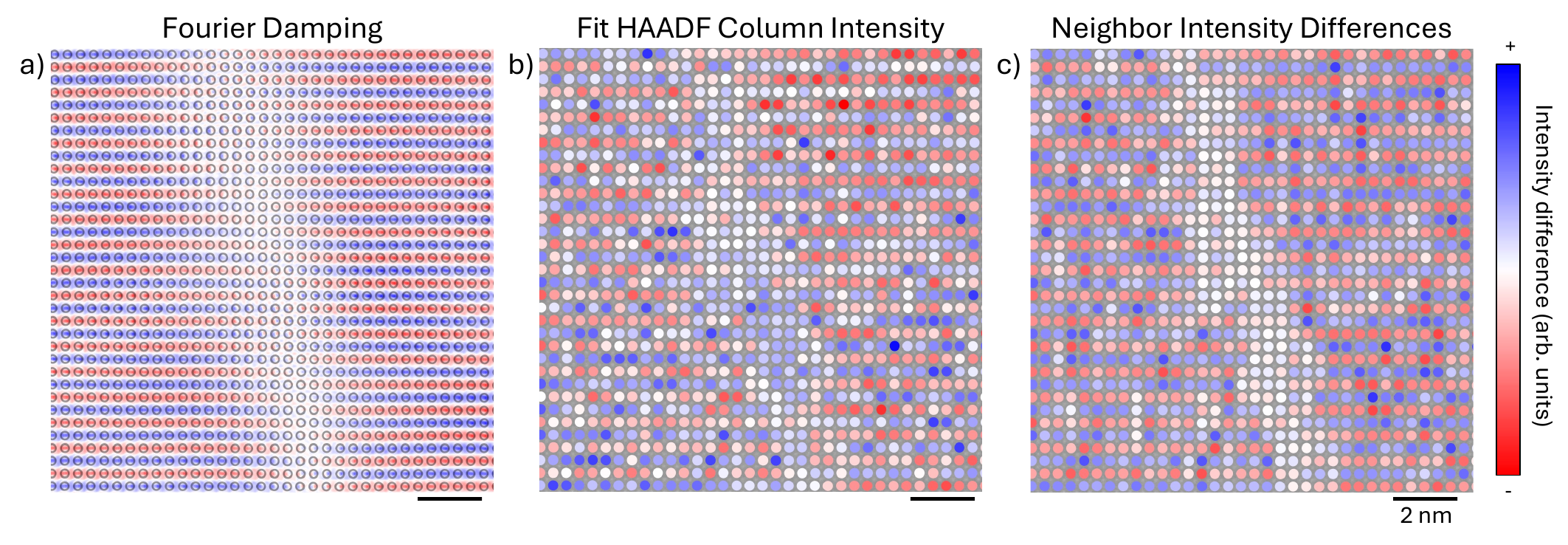}
    \caption{ \textcolor{black}{Comparison of defect characterization with Fourier damping analysis and direct Gaussian fitting of column intensities. 
    a) A-site column intensity variation measured with Fourier damping, where a diffuse anti-phase boundary extends vertically through the field of view, reproduced from Figure 3e. 
    b) A-site column intensities measured from direct fitting of the atomic columns in the HAADF image, where ordering is faintly visible but obscured by low frequency HAADF intensity variations and precision of fit intensities. 
    c) Directly fit A-site column intensities relative to the average of the neighboring columns above and below, reducing noise and emphasizing the A-site order, though a large variance in intensities remains. The diffuse anti-phase boundary is now clearly visible, with the same structure characterized by Fourier damping.
    }
    }
    \label{sfig:diffuse_boundary}
\end{figure*}

\begin{figure*}[ht]
    \includegraphics[clip=true,width=\textwidth]{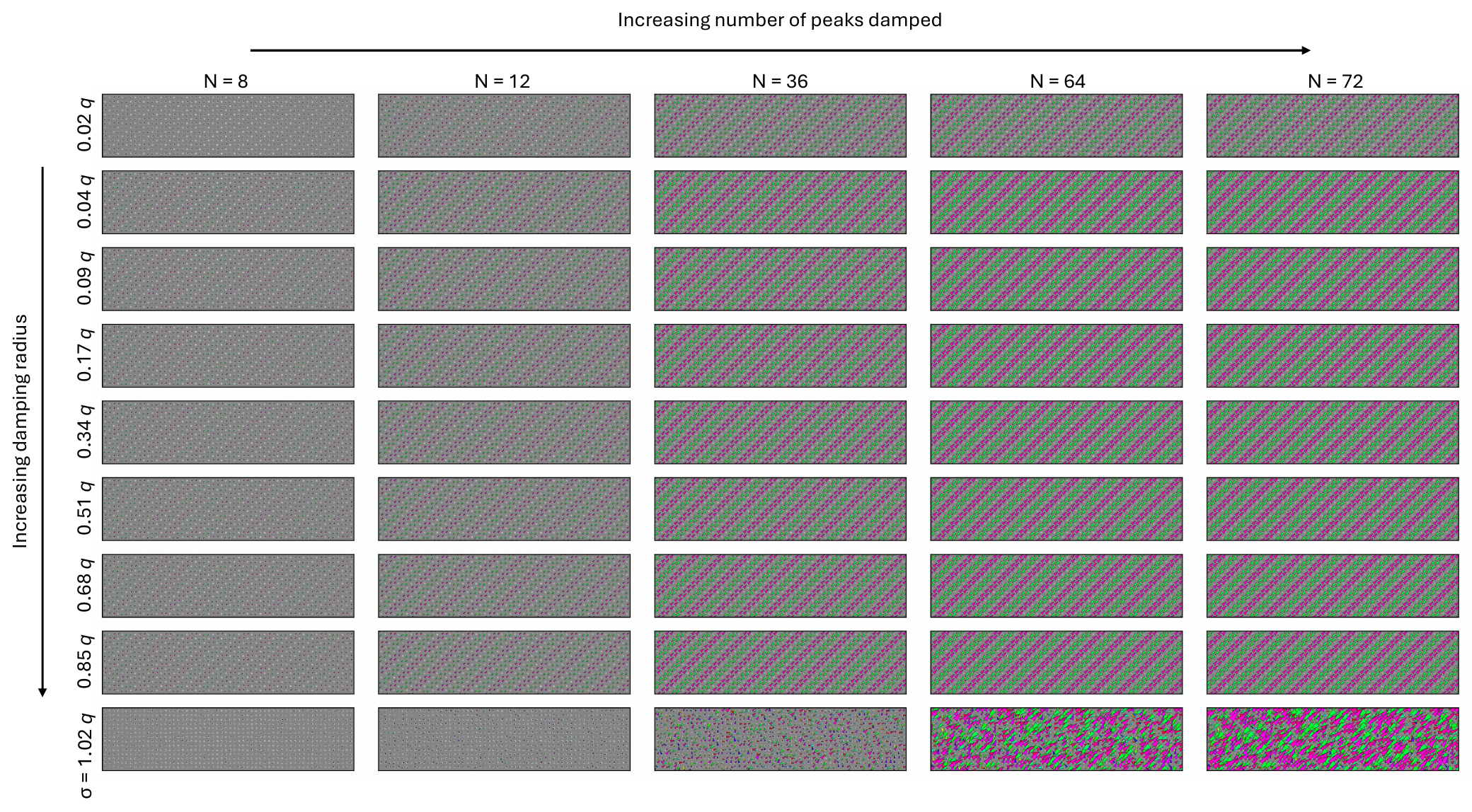}
    \caption{Maps of homogeneous distortion for various parameter choices.
    }
    \label{sfig:sim_homog_maps}
\end{figure*}

\begin{figure*}[ht]
    \includegraphics[clip=true,width=\textwidth]{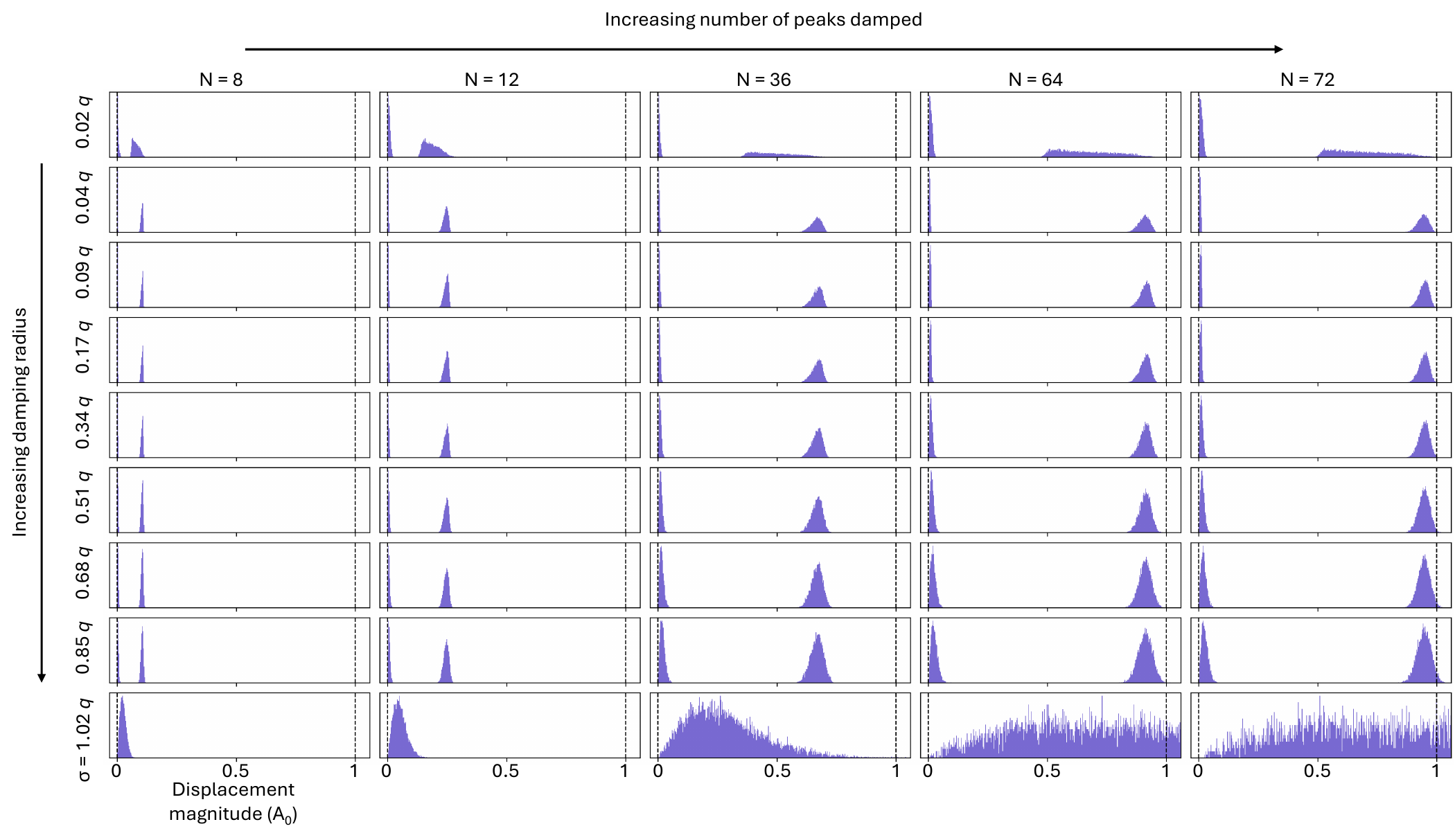}
    \caption{Histograms of homogeneous distortion magnitudes for various parameter choices. Black dashed lines at 0 and A$_0$ indicate the true displacement magnitudes in the modeled structure.
    }
    \label{sfig:sim_homog_hists}
\end{figure*}

\begin{figure*}[ht]
    \includegraphics[clip=true,width=\textwidth]{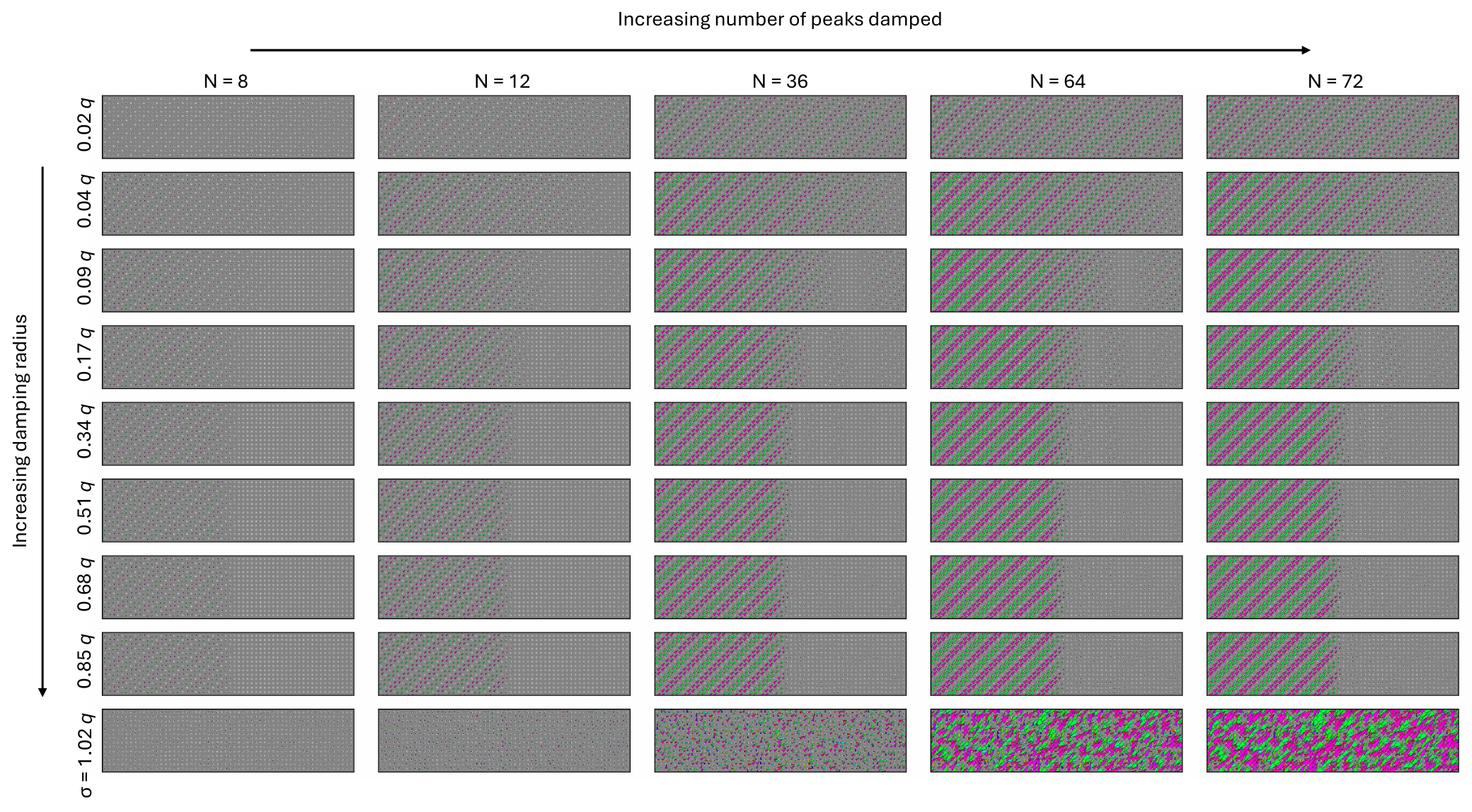}
    \caption{Maps of inhomogeneous distortion for various parameter choices.
    }
    \label{sfig:sim_inhomog_maps}
\end{figure*}

\begin{figure*}[ht]
    \includegraphics[clip=true,width=\textwidth]{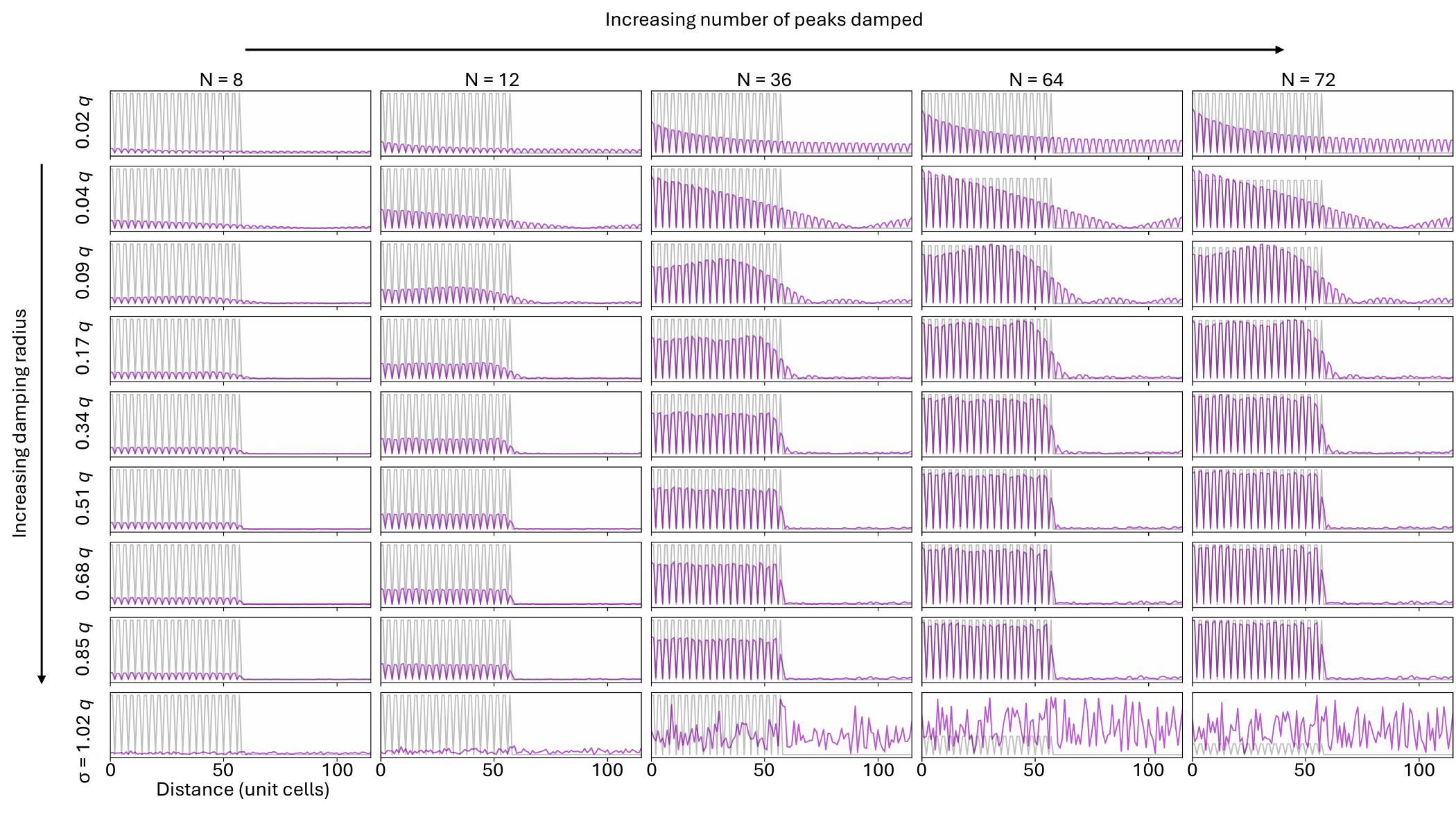}
    \caption{Displacement magnitude profiles of inhomogenous distortion for various parameter choices. Purple lines show magnitudes measured with various parameter pairs, gray lines show the true modeled displacement magnitudes. 
    }
    \label{sfig:sim_inhomog_profiles}
\end{figure*}

\begin{figure*}[ht]
    \includegraphics[clip=true,width=\textwidth]{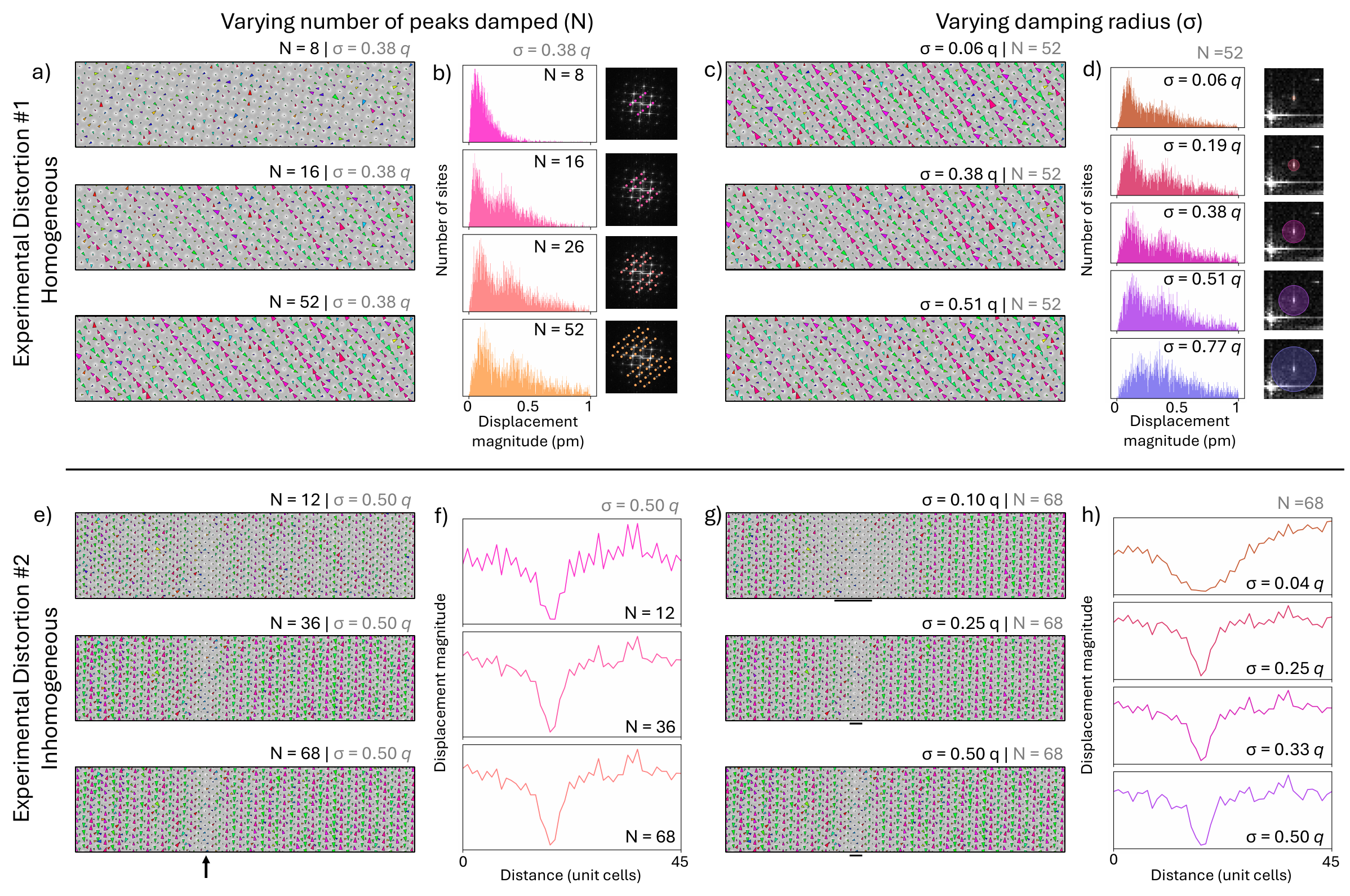}
    \caption{
    \textcolor{black}{
    Assessing Fourier damping parameter choices on experimental datasets. 
    a) Displacements measured on the same image of [001] oriented SmBaMn$_2$O$_6$ shown in Figure 2 for a series of different numbers of peaks damped. 
    b) Histograms (left) of displacement magnitudes showing increasing measured displacement amplitudes as the number of peaks damped is increased. The limited precision of the experimental STEM image results in broad distributions of measured displacements. Fourier transforms (right) show the peaks damped in each case.
    c) Maps of measured displacements on the same image for increasing radii damped around each peak from the top to bottom panels, while the number of peaks masked is held constant. 
    d) Histograms (left) showing the measured displacement magnitudes for various radii damped around the superlattice peaks. Fourier transforms cropped to center on a superlattice peak show the radii used in each case. Little response to the damping radius is seen unless the radius is so small the peak is not fully damped or so large that the tails of the neighboring Bragg peak are included.
    e) Displacements measured on an image of (La,Ca)MnO$_3$ (\cite{goodge2022disentangling} )from an antipolar distortion, in which an antiphase boundary extends vertically through the field of view (marked with black arrow), mapped for a series of increasing numbers of peaks damped.
    f) Profiles of displacement magnitudes across the interface, with a dip in displacement magnitude at the antiphase boundary, for varying numbers of peaks damped. As with the simulated datasets, the number of peaks damped does not significantly affect the measured sharpness of the  defect.
    g) Maps of measured displacements of the same image of  (La,Ca)MnO$_3$ for increasing radii damped around each peak showing increasing sharpness of the defect. Lines below the defect serve as a guide of its width for the eye.
    h) Profiles of displacement magnitudes across the interface for various damping radii show a large defect width at a very small damping radius (top), which quickly converges to a narrow width as the radius becomes large enough to fully include the peak.
    }
    }
    \label{sfig:exp_PLD_params}
\end{figure*}

\begin{figure*}[ht]
    \includegraphics[clip=true,width=\textwidth]{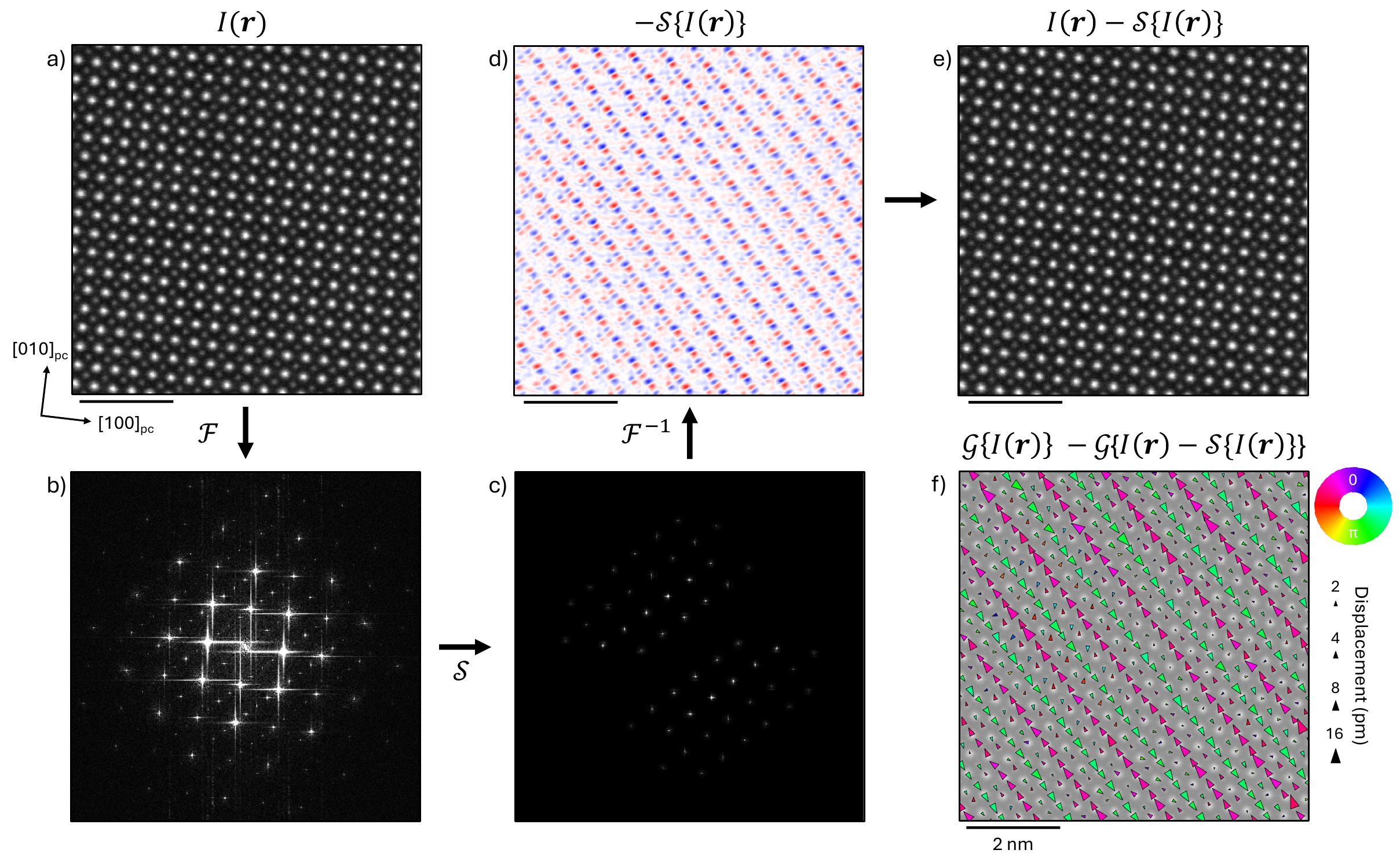}
    \caption{\textcolor{black}{
    Comparison of Fourier filtering to isolate symmetry breaking distortions.
    a) The same HAADF-STEM image shown in Figure 2a, and 
    b) its FFT. 
    c) Selecting only the superlattice peaks in the Fourier transform with identical Gaussian weightings with standard deviations set to half the circular damping radius used in Figure 2.
    d) The inverse Fourier transform of the superlattice peak signal, which closely resembles the difference image shown in Figure 2e with an added sign flip. 
    e) The difference between the original image and this Fourier filtered signal, which recovers a similar image to the undistorted reference signal shown in Figure 2d.
    f) Measured displacement vectors between the Gaussian fit atomic column positions in the original image (a) and the reference signal (e) yields a similar result to Figure 2f. 
    }}
    \label{sfig:filter_compare}
\end{figure*}

\begin{figure*}[ht]
    \includegraphics[clip=true,width=\textwidth]{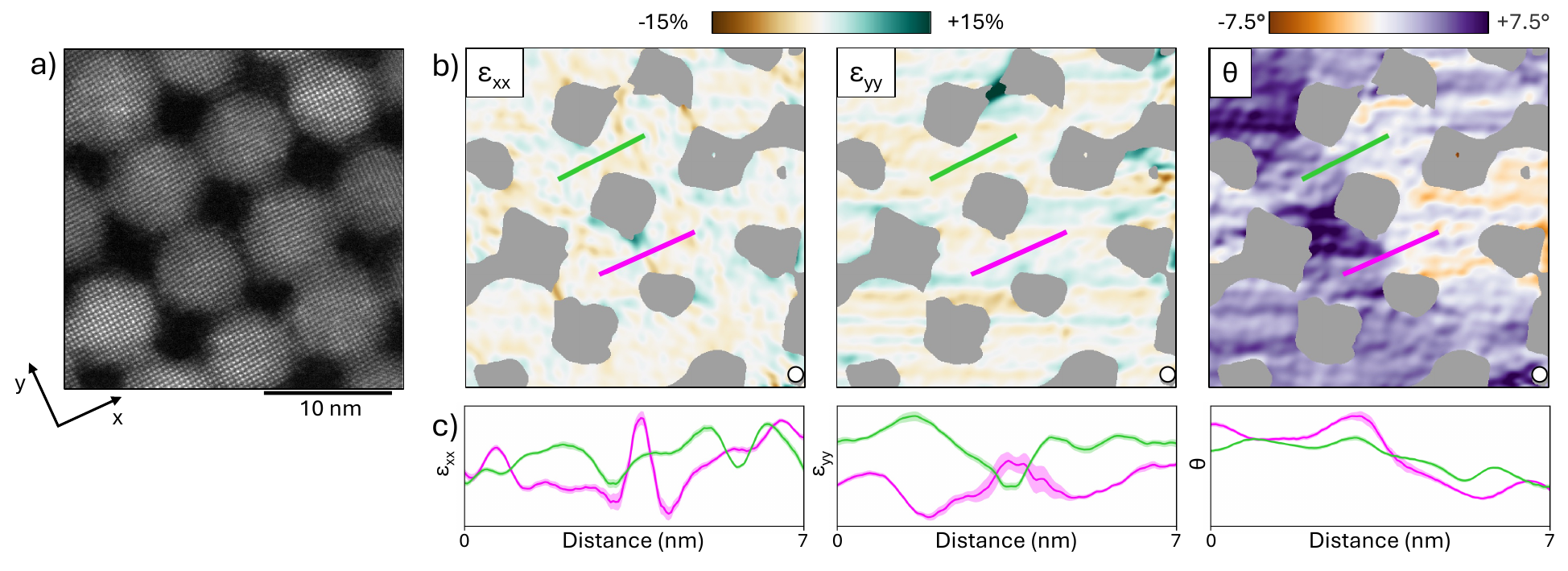}
    \caption{Measuring uncertainty in local wave fitting from fit variance.
    a) The HAADF-STEM image analyzed, in which one neck between a pair of quantum dots is continuous and the other shows a slight discontinuity.
    b) Maps of the measured strain and rotation, masked with the fit uncertainty, with profiles along the continuous (green) and discontinuous (pink) necks marked.
    c) Profiles of the measured strains and rotations (solid lines), with shaded areas surrounding the lines indicating the fit variance of the parameter plotted.
    }
    \label{sfig:wf_errorbars}
\end{figure*}